\documentclass[twocolumns]{aa} % for a paper on 1 column  
\usepackage{graphicx}
\usepackage{float}
\usepackage{subfig}
\usepackage{txfonts}
\usepackage{comment}
\usepackage{url}
\usepackage[switch]{lineno}
\usepackage{datetime}
\usepackage[usenames,table,dvipsnames]{xcolor}
\usepackage{setspace}
\usepackage[normalem]{ulem}
\usepackage{soul}
\definecolor{blue}{RGB}{66, 153, 233}
\definecolor{red}{RGB}{255, 0, 0}
\definecolor{purple}{RGB}{255, 0, 255}
\newcount\Comments  % 0 suppresses notes to selves in text
\newcommand{\kibitz}[2]{\ifnum\Comments=1\textcolor{#1}{#2}\fi}

\begin{document}

   \title{Three-dimensional morphological asymmetries in the ejecta of Cassiopeia A using a component separation method in X-rays}
   \titlerunning{Morphological asymmetries in the ejecta of Cassiopeia A in X-rays}
   \authorrunning{Picquenot et al.}
   \author{A. Picquenot$^1$
          \and
          F. Acero$^1$
          \and
          T. Holland-Ashford$^{2,3}$
          \and
          L.~A. Lopez$^{2,3}$
          \and
          J. Bobin$^1$
          }
 
          %\fnmsep\thanks{Just to show the usage
          %of the elements in the author field}

   \institute{\inst{1}~AIM, CEA, CNRS, Universit\'e Paris-Saclay, Universit\'e Paris Diderot, Sorbonne Paris Cit\'e, F-91191 Gif-sur-Yvette, France \\
   \inst{2}~Department of Astronomy, The Ohio State University, 140 W. 18th Ave., Columbus, OH 43210, USA\\
   \inst{3}~Center for Cosmology and AstroParticle Physics, The Ohio State University, 191 W. Woodruff Ave., Columbus, OH 43210, USA \\
   }
   
             %\email{adrien.picquenot@cea.fr, fabio.acero@cea.fr}
             %}

   \date{\today}

% \abstract{}{}{}{}{} 
% 5 {} token are mandatory
 
  \abstract{Recent simulations have shown that asymmetries in the ejecta distribution of supernova remnants can still reflect asymmetries from the initial supernova explosion. Thus, their study provides a great means to test and constrain model predictions in relation to the distributions of heavy elements or the neutron star kicks, both of which are key to better understanding the explosion mechanisms in core-collapse supernovae.
  
  The use of a novel blind source separation method applied to the megasecond X-ray observations of the well-known Cassiopeia A supernova remnant has revealed maps of the distribution of the ejecta endowed with an unprecedented level of detail and clearly separated from continuum emission. Our method also provides a three-dimensional view of the ejecta by disentangling the red- and blue-shifted spectral components and associated images of the Si, S, Ar, Ca and Fe, providing insights into the morphology of the ejecta distribution in Cassiopeia A. These mappings allow us to thoroughly investigate the asymmetries in the heavy elements distribution and probe simulation predictions about the neutron star kicks and the relative asymmetries between the different elements.
  
  We find in our study that most of the ejecta X-ray flux stems from the red-shifted component, suggesting an asymmetry in the explosion. In addition, the red-shifted ejecta can physically be described as a broad, relatively symmetric plume, whereas the blue-shifted ejecta is more similar to a dense knot. The neutron star also moves directly opposite to the red-shifted parts of the ejecta similar to what is seen with $^{44}$Ti. Regarding the morphological asymmetries, it appears that heavier elements have more asymmetrical distributions, which confirms predictions made by simulations. 
  This study is a showcase of the capacities of new analysis methods to revisit archival observations to fully exploit their scientific content.}
   %leave it empty if necessary  
   


   \maketitle
%
%-------------------------------------------------------------------

\section{Introduction}

Cassiopeia A (hereafter, Cas~A) is among the most studied astronomical objects in X-rays and is arguably the best-studied supernova remnant (SNR). Investigation of the distribution of metals on sub-parsec scales is possible because it is the youngest core-collapse (CC) SNR in the Milky Way \citep[about $340$ years old;][]{thorstensen01}, its X-ray emission is dominated by the ejecta metals \citep{hwang12}, and it is relatively close \citep[$3.4$ kpc, see][]{1995ApJ...440..706R,2014MNRAS.441.2996A}. Cas~A benefits from extensive observations (about $3$ Ms in total by {\it Chandra}), making it an ideal laboratory to probe simulation predictions regarding the distribution of ejecta metals.

In the last few years, three-dimensional simulations of CC supernovae (SNe) have begun to produce testable predictions of SNe explosion and compact object properties in models using the neutrino-driven mechanism \citep[see][]{janka16,muller16,Burrows_2019,Burrows_Vartanyan20}. In particular, explosion-generated ejecta asymmetries \citep{wongwathanarat13,summa18,janka17} and neutron star (NS) kick velocities \citep{delaney13} appear to be key elements in CC SN simulations that Cas~A's data can constrain. Although it is challenging to disentangle the asymmetries produced by the surrounding medium from those inherent to the explosion, \cite{Orlando2016} have explored the evolution of the asymmetries in Cas~A using simulations beginning from the immediate aftermath of the SN and including the three-dimensional interactions of the remnant with the interstellar medium. Similar simulations presenting the evolution of a Type Ia SNR over a period spanning from one year after the explosion to several centuries afterward have been made by \cite{Ferrand_2019}, showing that asymmetries present in the original SN can still be observed after centuries. The same may go for the CC SNR Cas~A, and a better knowledge of its three-dimensional morphology could lead to a better understanding of the explosion mechanisms by providing a way to test the simulations.

An accurate mapping of the different elements' distributions, the quantification of their relative asymmetries, and their relation to the NS motion would, for example, allow us to probe the simulation predictions that heavier elements are ejected more asymmetrically and more directly opposed to the NS motion than lighter elements \citep{wongwathanarat13,janka17,Gessner_2018,M_ller_2019}. On this topic, this paper can be viewed as a follow-up to \cite{hollandashford20}, a study that aimed to quantitatively compare the relative asymmetries of different elements within Cas~A, but which was hindered due to difficulties in separating and limiting contamination in the elements' distribution. Moreover, in that analysis, the separation of the blue- and red-shifted parts in these distributions was not possible.

Here, we intend to fix these issues by using a new method to retrieve accurate maps for each element's distribution, allowing us to further investigate their individual and relative physical properties. This method is based on the General Morphological Components Analysis \citep[GMCA, see][]{bobin15}, a blind source separation (BSS) algorithm that was introduced for X-ray observations by \cite{picquenot:hal-02160434}. It can disentangle both spectrally and spatially mixed components from an X-ray data cube of the form $(x,y,E)$ with a precision unprecedented in this field. The new images thus obtained suffer less contamination by other components, including the synchrotron emission. It also offers the opportunity to separate the blue- and red-shifted parts of the elements' distribution, thereby facilitating a three-dimensional mapping of the X-ray emitting metals and a comparison of their relative asymmetries. Specifically, the GMCA is able to disentangle detailed maps of a red- and a blue-shifted parts in the distributions of Si, S, Ca, Ar, and Fe, thus providing new and crucial information about the three-dimensional morphology of Cas~A. This is a step forward as previous studies intending to map the distribution of the individual elements and study their asymmetries in Cas~A in X-rays \citep{hwang12,katsuda18,hollandashford20} were not able to separate red- and blue-shifted components.

This paper is structured as follows. In Section~\ref{sect:method}, we will describe the nature of the data we use (Section~\ref{sub:data}), our extraction method (Section~\ref{sub:extraction}), our way to quantify the asymmetries (Section~\ref{sub:asym_q}), and our method to retrieve error bars (Section~\ref{sub:error}). In Section~\ref{sect:method}, we will present the images resulting from the application of our extraction method (Sections~\ref{sub:image_1} and~\ref{sub:image_2}), and we will discuss the interpretation of the retrieved images as blue- or red-shifted by looking at their associated spectra (Section~\ref{sub:spectra}), and will present the results of a spectral analysis on these same spectra (Section~\ref{sub:xspec}). Lastly, we will discuss in Section~\ref{sect:physics} the physical information we can infer from our results. Section~\ref{sub:int_prm} will be dedicated to the interpretation of the spatial asymmetries of each line emission, while Sections~\ref{sub:int_velocity} and \ref{sub:int_NSvelocity} will focus respectively on the mean direction of each line's emission and on the NS velocity. A comparison with the {\it NuSTAR} data of $^{44}$Ti will finally be presented in Section~\ref{sect:ti}.

%--------------------------------------------------------------------
\section{Method}
\label{sect:method}
\subsection{Nature of the data}
\label{sub:data}

Spectro-imaging instruments, such as those aboard the current generation of X-ray satellites {\it XMM-Newton} and {\it Chandra}, provide data comprising spatial and spectral information: The detectors record the position $(x,y)$ and energy $E$ event by event, thereby producing a data cube with two spatial dimensions and one spectral dimension. For our study, we used {\it Chandra} observations of the Cas~A SNR, which was observed with the ACIS-S instrument in 2004 for a total of 980 ks \citep{hwang04}. We used only the 2004 data set to avoid the need to correct for proper motion across epochs. The event lists from all observations were merged in a single data cube. 
The spatial (of 2\arcsec) and spectral binning (of 14.6~eV) were adapted so as to obtain a sufficient number of counts in each cube element. No background subtraction or vignetting correction has been applied to the data.

\subsection{Image extraction}
\label{sub:extraction}

%In order to study asymmetries in the ejecta metals in Cas~A, a good mapping of their spatial distribution is needed. However, extracting the spatial distribution of each element is not a straightforward process as multiple components, such as the shocked ejecta and the synchrotron emission, are overlapping, sometimes with a high contrast factor. \citet{picquenot:hal-02160434} introduced a method that was able to disentangle both morphologically- and spectrally-accurate components from a $(x,y,E)$ X-ray data cube. This method was based on the GMCA, a BSS algorithm first introduced in \citet{bobin15}. 

The main concept of GMCA is to take into account the morphological particularities of each component in the wavelet domain to disentangle them, without any prior instrumental or physical information. Apart from the $(x,y,E)$ data cube, the only input needed is the number $n$ of components to retrieve, which is user-defined. The outputs are then a set of $n$ images associated with $n$ spectra. Each couple image-spectrum represents a component: The algorithm makes the assumption that every component can be described as the product of an image with a spectrum. Thus, the retrieved components are approximations of the actual components with the same spectrum on each point of the image. Nevertheless, \citet{picquenot:hal-02160434} showed that when tested on Cas~A-like toy models, the GMCA was able to extract morphologically and spectrally accurate results. The tested spectral toy models included power-laws, thermal plasmas, and Gaussian lines. In particular, in one of these toy models, the method was able to separate three components: two nearby partially overlapping Gaussian emission lines and power-law emission. The energy centroids of both Gaussians were accurately retrieved, despite their closeness. Such a disentangling of mixed components with similar neighboring spectra cannot be obtained through line-interpolation, and fitting of a two-Gaussian model region by region is often time consuming, producing images contaminated by other components with unstable fitting results. 

In the same paper, the first applications on real data of Cas~A were promising, in particular concerning asymmetries in the elements' distribution. For Si, S, Ar, Fe and Ca, the GMCA was able to retrieve two maps associated with spectra slightly blue- or red-shifted from their theoretical position. The existence of blue- or red-shifted parts in these elements' distribution was previously known, and the Fe maps from \citet{picquenot:hal-02160434} were consistent with prior works but endowed with more details \citep[see][]{2002A&A...381.1039W,delaney10}. Thus, they constitute a great basis for an extensive study of the asymmetries in the elements' distribution in Cas~A.

In this paper, we will use a more recent version of the GMCA, the pGMCA, that was developed to take into account data of a Poissonian nature \citep[][]{9215040}. In the precedent version of the algorithm, the noise was supposed to be Gaussian. Even with that biased assumption, the results were proven to be reliable. However, a proper treatment of the noise is still relevant: It increases the consistency of the spectral morphologies of the retrieved components and makes the algorithm able to disentangle components with a fainter contrast.

The mathematical formalism is similar to that of the GMCA, presented in \citet{picquenot:hal-02160434}. The fundamental difference is that instead of a linear representation, the pGMCA uses the notion of a Poisson-likelihood of a given sum of components to be the origin of a certain observation. The problem solved by the algorithm is thus essentially the same kind, the main difference being a change in the nature of
the norm that needs to be minimized. A more precise description of this new method is available in \cite{9215040}.

The use of the pGMCA is also highly similar to that of the GMCA. One notable difference is that the pGMCA is more sensitive to the initial conditions, so it needs a first guess for convergence purposes. The analysis therefore consists of two steps: a first guess obtained with the GMCA and a refinement step using the Poissonian version pGMCA.

\begin{figure}
\centering
\includegraphics[width=9cm]{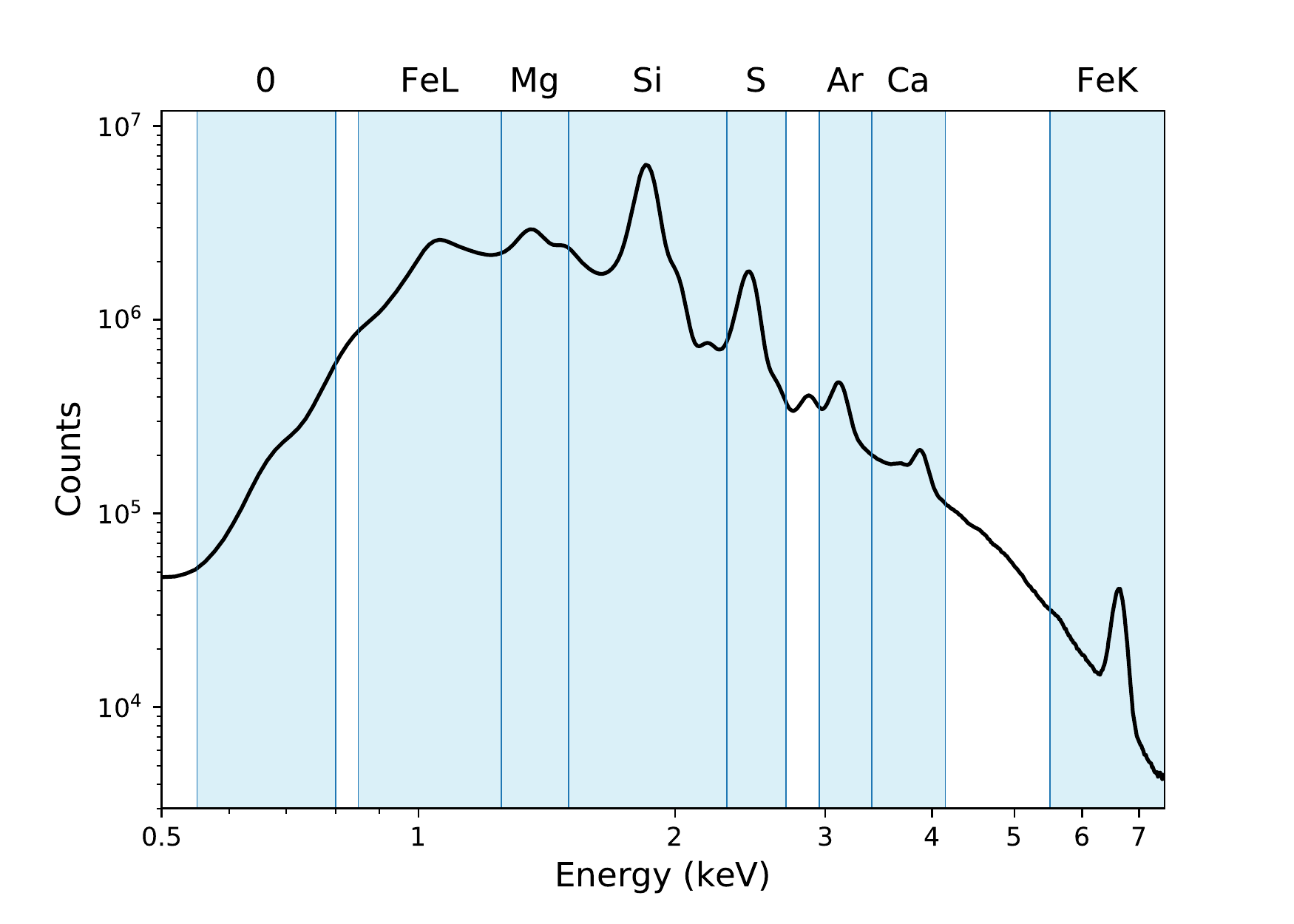}
\caption{Spectrum of Cas~A obtained from the combination of the deep {\it Chandra} 2004 observations. The source separation algorithm was applied in each individual energy band band, which are represented by the shaded regions.}
\label{fig:E-bands}
\end{figure}

The aforementioned workflow was applied to the Cas~A {\it Chandra} observations by creating data cubes for each energy band shown in Fig.~\ref{fig:E-bands}. These energy bands were chosen to be large enough to have the leverage to allow the synchrotron continuum to be correctly retrieved and to be narrow enough to avoid contamination by other line emissions. The pGMCA being a fast-running algorithm, the final energy bands were chosen after tests to find the best candidates for both criteria. For each band, the initial number of components $n$ was $3$: the synchrotron emission and the blue- and red-shifted parts of the line emission. We then tested using $4$ and $5$ components to ensure extra components were not merged into our components of interest. We also tested with $2$ components to verify our assumption on the presence of blue- and red-shifted parts was not imposing the apparition of a spurious component. For each emission line, we then chose $n$ as the best candidate to retrieve the most seemingly meaningful components without spurious images.

For each analysis, the algorithm was able to retrieve a component that we identify as the synchrotron emission (a power-law spectrum and filamentary spatial distribution, not shown here) and multiple additional thermal components with strong line features. We were able to identify two associated images with shifted spectra from the theoretical emission line energy for all these line features except O,  Fe~L, and Mg.

%\textbf{Notes:} 
%\begin{itemize}
%    \item In some energy bands, we might be recovering different ions emission like Si XIII and SI XIV probably associated with components revealing a plasma of different temperature/abundance/ionization time. Do we present it here, in Appendix or not at all (different study) ?
 %   \item 
%\end{itemize}

\subsection{Quantification of asymmetries}
\label{sub:asym_q}

We use the power-ratio method (PRM) to quantitatively analyze and compare the asymmetries of the images extracted by pGMCA. This method was developed by \citet{1995ApJ...452..522B} and previously employed for use on SNRs \citep{lopez09b,Lopez_2009,lopez11}. It consists of calculating multipole moments in a circular aperture positioned on the centroid of the image, with a radius that encloses the whole SNR. Powers of the multipole expansion $P_m$ are then obtained by integrating the $m$th term over the circle. To normalize the powers with respect to flux, they are divided by $P_0$, thus forming the power ratios $P_{m}/P_0$. For a more detailed description of the method, see \citet{Lopez_2009}.

The $P_2/P_0$ and $P_3/P_0$ terms convey complementary information about the asymmetries in an image. The first term is the quadrupole power-ratio and quantifies the ellipticity/elongation of an extended source, while the second term is the octupole power-ratio and is a measure of mirror asymmetry. Hence, both are to be compared simultaneously to ascertain the asymmetries in different images.

Here, as we want to compare asymmetries in the blue- and red-shifted part of the elements' distribution, the method is slightly modified. In a first step, we calculate the $P_2/P_0$ and $P_3/P_0$ ratios of each element's total distribution by using the sum of the blue- and red-shifted maps as an image. Its centroid is then an approximation of the center-of-emission of the considered element. Then, we calculate the power ratios of the blue- and red-shifted images separately using the same center-of-emission. Ultimately, we normalize the power ratios thus obtained by the power ratios of the total element's distribution:

\begin{equation}
P_i/P_{0\textit{  (shifted / total)}} =\frac{P_i/P_{0\textit{  (red or blue image)}}}{P_i/P_{0\textit{  (total image)}}},
\end{equation}
\label{eq:Pshifted}

\noindent
where $i=2$ or $3$ and $P_i/P_{0\mbox{ (red or blue image)}}$ is calculated using the centroid of the total image. That way, we can compare the relative asymmetries of the blue- and red-shifted parts of different elements, without the comparison being biased by the original asymmetries of the whole distribution.

\subsection{Error bars}
\label{sub:error}

As explained in \citet{picquenot:hal-02160434}, error bars can be obtained by applying this method on every image retrieved by the GMCA applied on a block bootstrap resampling. However, as was shown in that paper, this method introduces a bias in the results of the GMCA. We show in Appendix~\ref{ap:boot} that the block bootstrap method modifies the Poissonian nature of the data, thus impacting the results of the algorithm. Since the pGMCA is more dependent than GMCA on the initial conditions, the bias in the outputs is even greater with this newer version of the algorithm (see Fig.~\ref{fig:bias_sync}). For that reason, we developed a new resampling method we named "constrained bootstrap," presented in Appendix~\ref{sec:newboot}.

Thus, we applied pGMCA on a hundred resampled data cubes obtained thanks to the constrained bootstrap and plotted the different spectra we retrieved around the ones obtained on real data. As stated in Appendix~\ref{sec:newboot}, the spread between the resamplings has no physical significance but helps in evaluating the robustness of the algorithm around a given set of original conditions. The blue-shifted part of the Ca line emission, a very weak component, was not retrieved for every resampling. In this case, we created more resamplings in order to obtain a hundred correctly retrieved components. The faintest components are the ones with the largest relative error bars, as can be seen in Fig.~\ref{fig:red_blue_images} and Fig.~ \ref{fig:other_images}, highlighting the difficulty for the algorithm to retrieve them in a consistent way on a hundred slightly different resamplings.

To obtain the error bars for the PRM plot of the asymmetries, we  applied the PRM to the hundred images retrieved by the pGMCA on the resamplings. Then, in each direction we plotted error bars representing the interval between the $10^{\rm th}$ and the $90^{\rm th}$ percentile and crossing at the median. We also plotted the PRM applied on real data. Although our new constrained bootstrap method ensures the Poissonian nature of the data to be preserved in the resampled data sets, we see that the results of the pGMCA on real data are sometimes not in the $10^{\rm th}$-$90^{\rm th}$ percentile zone, thus suggesting there may still be some biases. It happens mostly with the weakest components, showing once more the difficulty for the pGMCA to retrieve them consistently out of different data sets presenting slightly different initial conditions. However, even when the results on real data are not exactly in the $10^{\rm th}$-$90^{\rm th}$ percentile zone, the adequation between the results on real and resampled data sets is still good, and the relative positioning for each line is the same, whether we consider the results on the original data or on the resampled data sets.

%However, this plot is not interesting for the values of the multipole moments in themselves, but for the comparison between the positioning of the different line emissions in the plot, and in spite of the possible biases, their relative positioning is the same.

%The lack of robustness of the pGMCA algorithm is a non-trivial problem that goes far beyond the scope of this paper. However, this is an issue that needs to be expressed, and we think that showing the errorbars obtained through block bootstrap resamplings is a way to highlight it.

\section{Results}
\label{sect:result}

\subsection{Images retrieved by pGMCA}
\label{sub:image_1}

\begin{figure}
\centering
\includegraphics[width=9cm]
{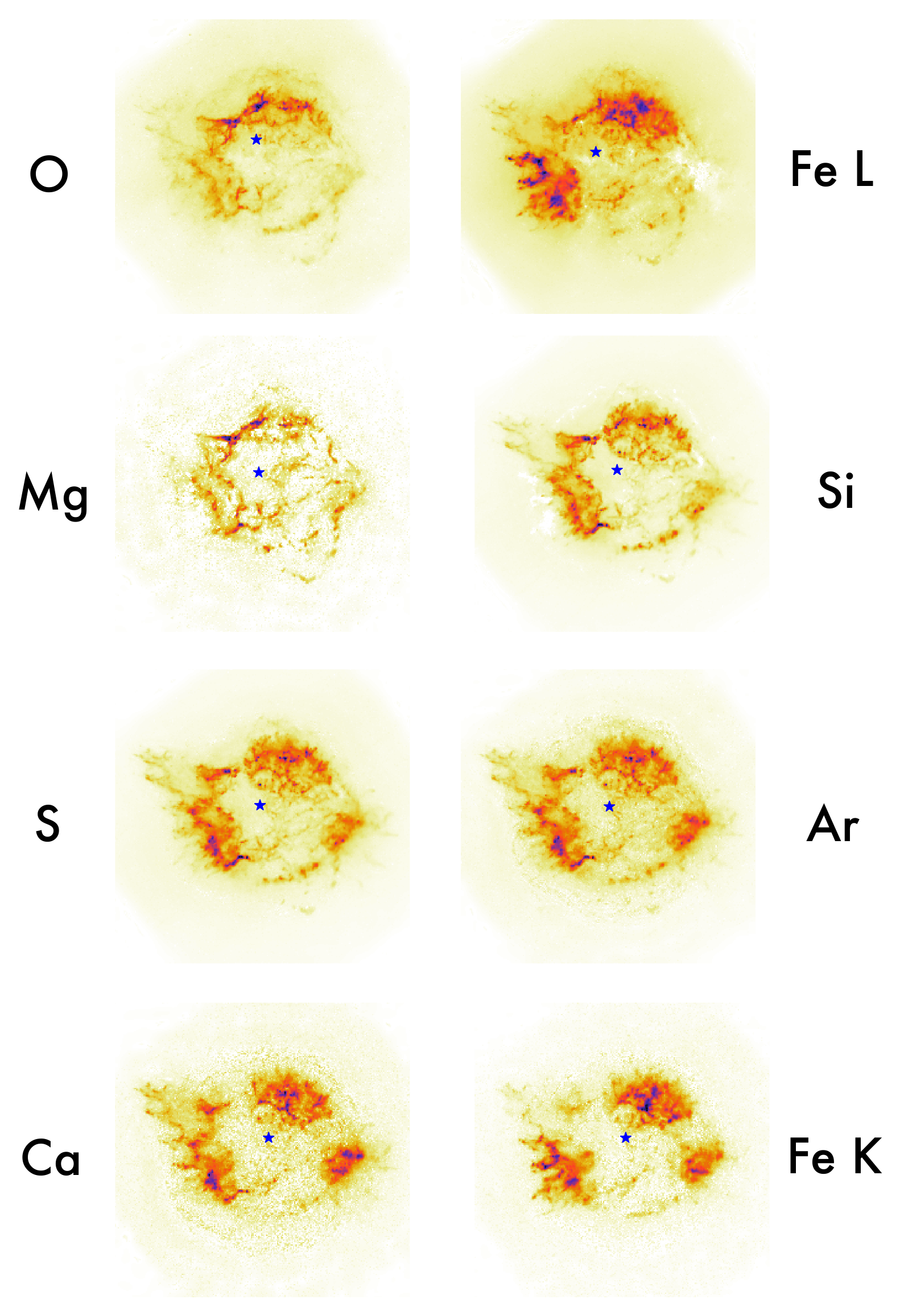}
\caption{Total images of the different line emission spatial structure as retrieved by the pGMCA. The blue symbol represents the image centroid adopted in the PRM analysis. The color-scale is in square root.}
\label{fig:images_tot}
\end{figure}

\begin{figure*}
\centering
\includegraphics[width=17.6cm]
{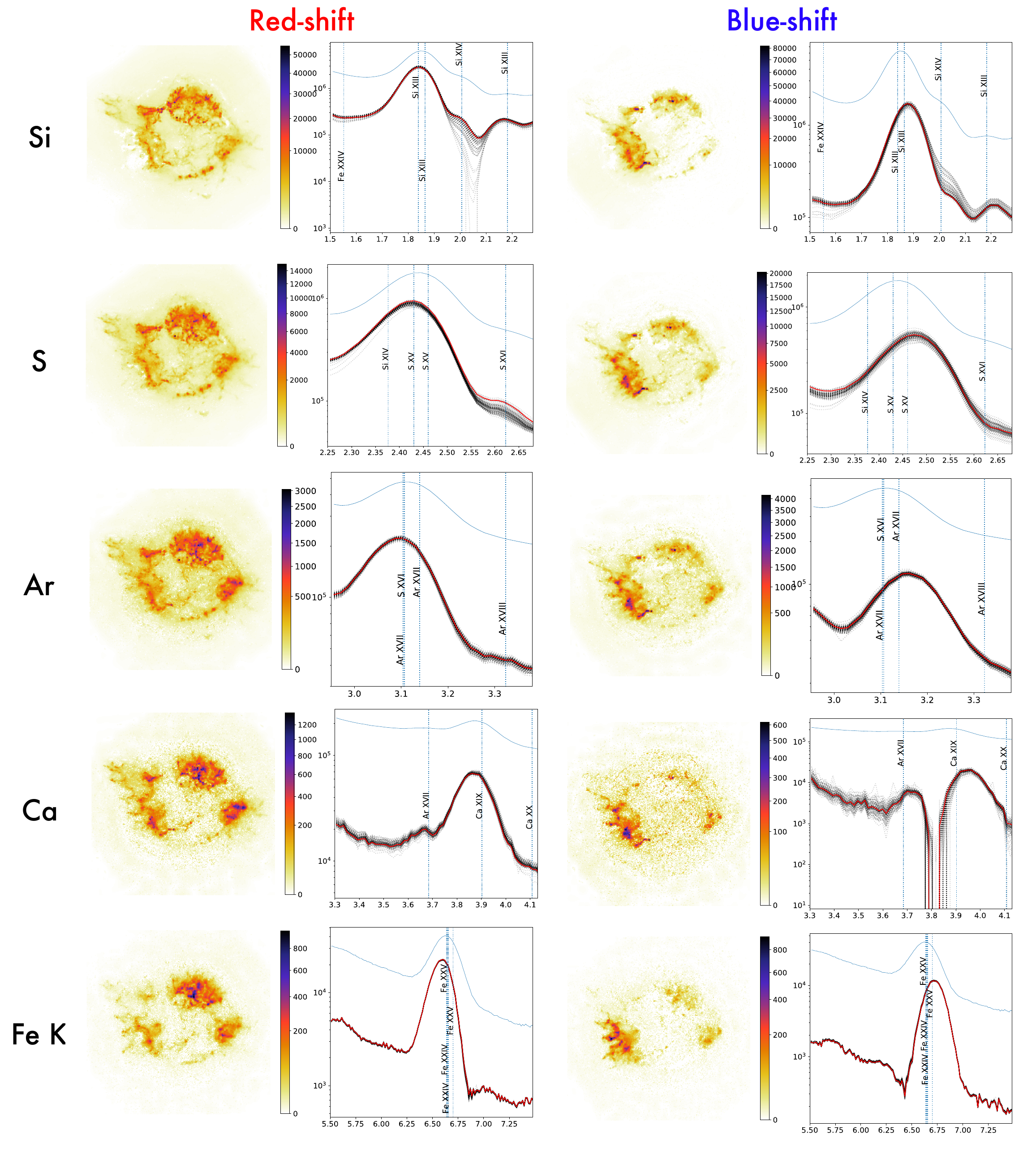}
\caption{Red- and blue-shifted parts of the Si, S, Ar, Ca, and Fe line emission spatial distribution and their associated spectrum as found by pGMCA. The spectra in red correspond to the application of the algorithm on real data, while the dotted gray spectra correspond to the application on a hundred constrained bootstrap resamplings illustrating statistical uncertainties. The x-axis is in keV and the y-axis in counts.
The dotted \textbf{vertical} lines represent the energy of the brightest emission lines for a non-equilibrium ionization plasma at a temperature of 1.5 keV and ionization timescale of $log(\tau)=11.3$ cm$^{-3}$ s produced using the AtomDB \citep{2012ApJ...756..128F}. These parameters are the mean value of the distribution shown in Fig.2 of \citet{hwang12}.}
\label{fig:red_blue_images}
\end{figure*}

\begin{figure}
\centering
\includegraphics[width=9cm]
{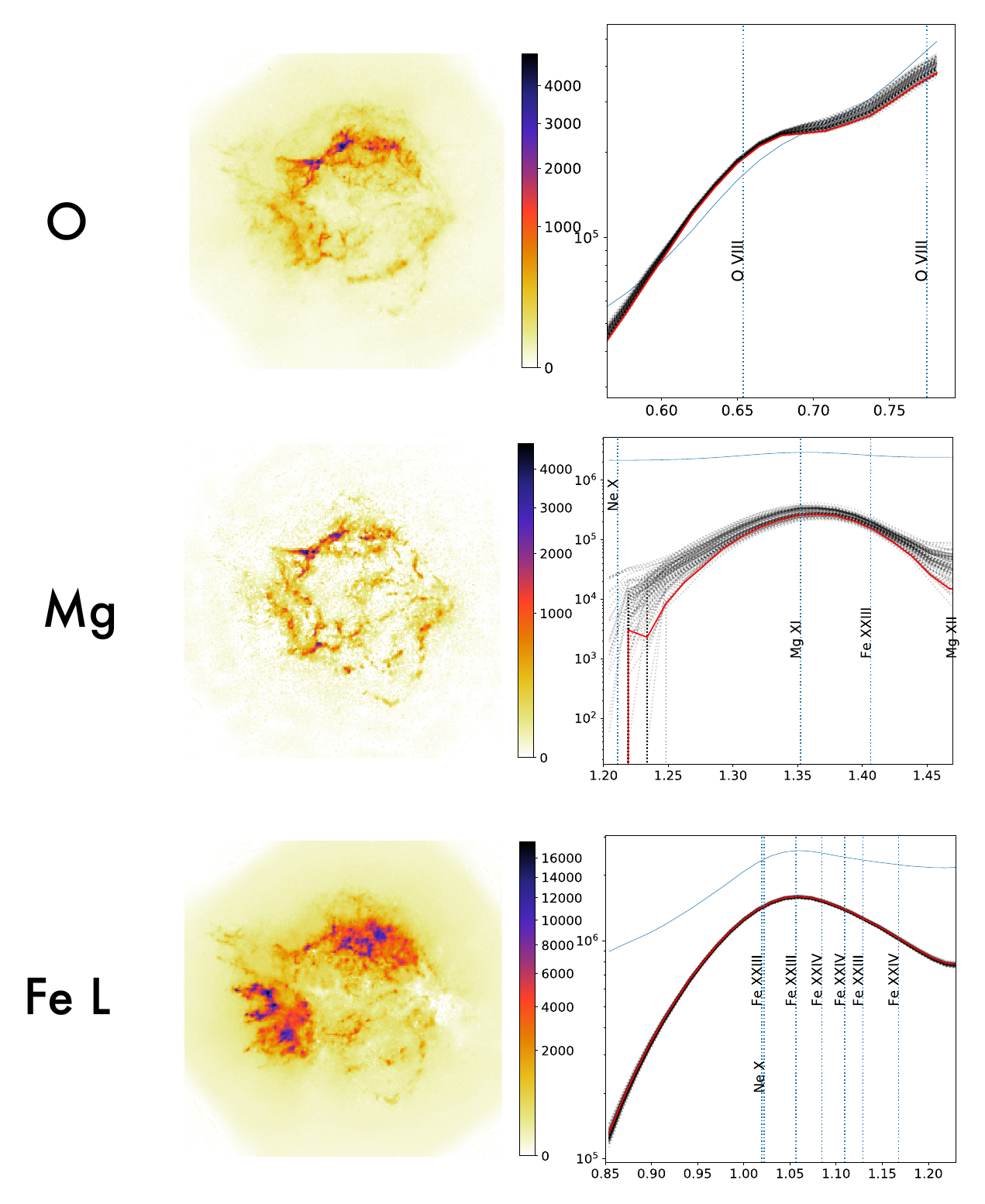}
\caption{Images of the O, Mg, and Fe L line emission spatial structures and their associated spectra as found by pGMCA. The spectra in red correspond to the application of the algorithm on real data, while the dotted gray spectra correspond to the application on a hundred constrained bootstrap resamplings. The x-axis is in keV and the y-axis in counts.}
\label{fig:other_images}
\end{figure}

By applying the pGMCA algorithm on the energy bands surrounding the eight emission lines shown in Fig.~\ref{fig:E-bands}, we were able to retrieve maps of their spatial distribution associated with spectra, successfully disentangling them from the synchrotron emission or other unwanted components. The O, Mg, and Fe L lines were only retrieved as single features, each associated with a spectrum, whereas Si, S, Ar, Ca, and Fe-K were retrieved as two different images associated with spectra that we interpret as being the same emission lines slightly red- or blue-shifted. Fig.~\ref{fig:images_tot} shows the total images for all eight line emissions, obtained by summing the blue- and red-shifted parts when necessary. It also indicates the centroid of each image that is adopted in the PRM. Fig.~\ref{fig:red_blue_images} shows the red- and blue-shifted parts of five line emissions, together with their associated spectra, while Fig.~\ref{fig:other_images} presents the images of O, Mg, and Fe L together with their respective spectra. Fig.~\ref{fig:images_tot} is similar to Fig. 10 from \cite{picquenot:hal-02160434}, but the images here are more accurate and less contaminated by other components thanks to a proper treatment of the Poisson noise, and the associated spectra not shown in our first paper are presented here in Fig.~\ref{fig:red_blue_images} and Fig.~\ref{fig:other_images}.

\subsection{Discussion on the retrieved images}
\label{sub:image_2}

The fact that our algorithm fails to separate a blue-shifted from a red-shifted part in the O, Mg, and Fe L images is not surprising. At $1$~keV, we infer that a radial speed of $4000$ km~s$^{-1}$ would lead to a $\Delta E$ of about $13$~eV, which is below the spectral bin size of our data. We see in Fig.~\ref{fig:images_tot} that while the O and the Mg images are highly similar, they are both noticeably different from the images of the other line emissions. Both the O and Mg images exhibit similar morphology to the optical images of O~{\sc ii} and O~{\sc iii} from Hubble \citep{2001AJ....122.2644F,Patnaude_2014}. The intermediate mass elements share interesting properties: Their spatial distributions appears similar in Fig.~\ref{fig:images_tot}, and the division into a red- and a blue-shifted part (as found by the pGMCA) allows us to investigate their three-dimensional morphology. We also notice that the maps of Si and Ar are similar to that of the Ar{\sc ii} in infrared \citep{delaney10}.

As the reverse shock has not fully propagated to the interior of Cas~A \citep{Gotthelf01,delaney10}, our images may not reflect the full distribution of the ejecta. However, \cite{hwang12} estimates that most of the ejecta mass has already been shocked: we can thus conclude that our images capture the bulk of the ejecta and our element images are likely similar to the true ejecta distributions.
In addition, our Fe red-shifted image matches well with the $^{44}$Ti image, produced by radioactive decay instead of reverse-shock heating (see Section~\ref{sect:ti}).

%\subsection{Discussion on red/blue images}
%
%\begin{itemize}
%
%    \item could fit Gaussian model for each line %and derive a velocity red/blue for each %element and plot Vejecta vs asym.
%    This is difficult as we don't have a good way %to estimate errors and hence to carry out a %good fit.   
%    \item Could the red/blue shifted vectors %inform us about the recoil of the NS, at %least its angle wrt LOS ? assuming ejecta %assym trace explosion assym.
%\end{itemize}

\begin{figure}
\centering
\includegraphics[width=9cm]
{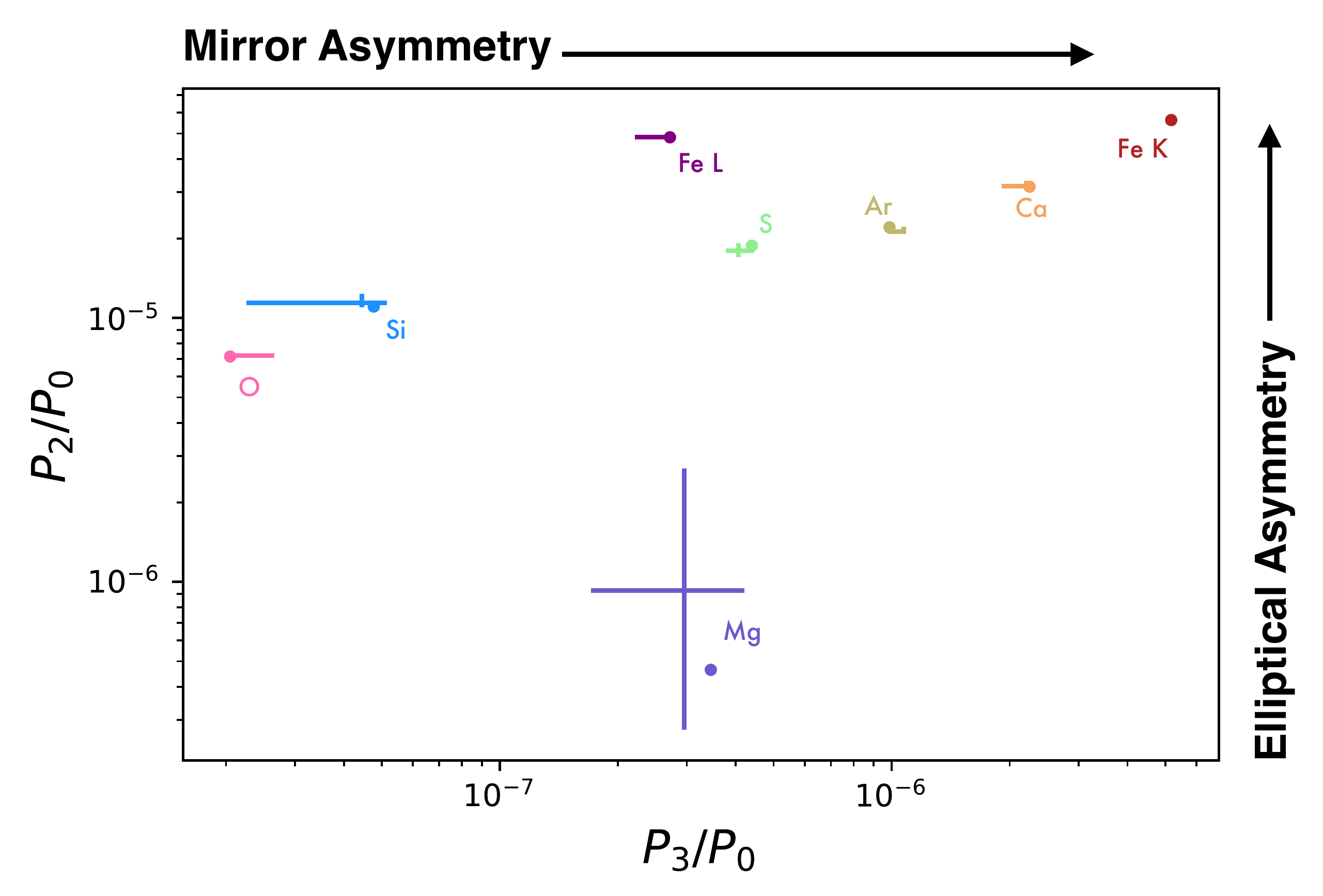}
\caption{Quadrupole power-ratios $P_{2}/P_{0}$ versus the octupole power-ratios $P_{3}/P_{0}$ of the total images of the different line emissions shown in Fig.~\ref{fig:images_tot}. The dots represent the values measured for the pGMCA images obtained from the real data, and the crosses the $10^{\rm th}$ and $90^{\rm th}$ percentiles obtained with pGMCA on a hundred constrained bootstrap resamplings, with the center of the cross being the median.} 
\label{fig:asym}
\end{figure}

\begin{figure}
\centering
\includegraphics[width=9cm]
{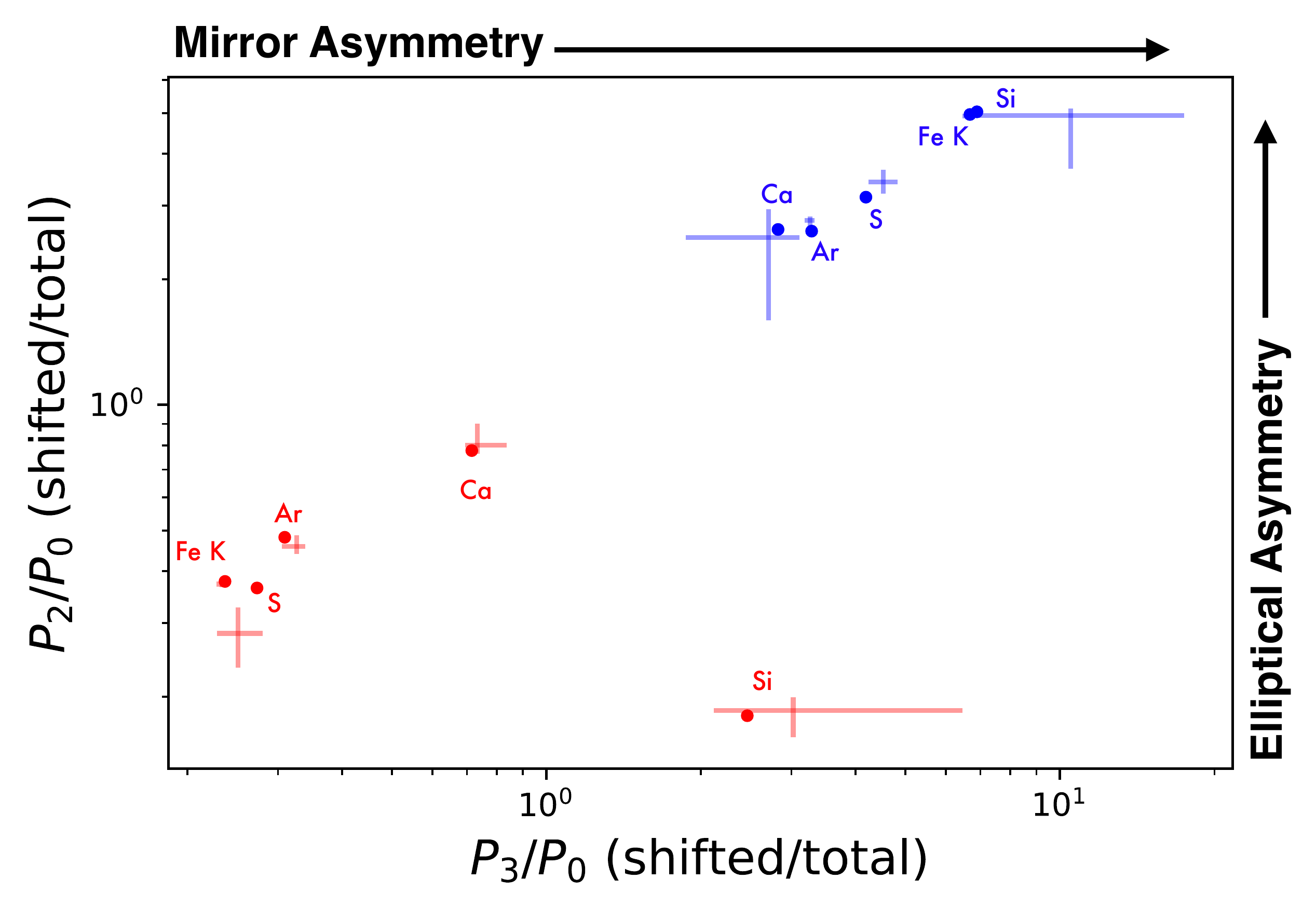}
\caption{Quadrupole power-ratios $P_{2}/P_{0}$ versus the octupole power-ratios $P_{3}/P_{0}$ of the red- and blue-shifted images of the different line emissions shown in Fig. \ref{fig:red_blue_images}, normalized with the quadrupole and octupole power-ratios of the total images. The dots and error bars are obtained in the same way as in Fig.~\ref{fig:asym}}
\label{fig:red_blue}
\end{figure}

Hence, we can quantify the asymmetries in the ejecta distribution by using the PRM method described in Sect.~\ref{sub:asym_q} on our images. Fig.~\ref{fig:asym} presents the quadrupole power-ratios $P_{2}/P_{0}$ versus the octupole power-ratios $P_{3}/P_{0}$ of the total images from Fig.~\ref{fig:images_tot}. Fig.~\ref{fig:red_blue} shows the quadrupole power-ratios versus the octupole power-ratios of the red- and blue-shifted images presented in Fig.~\ref{fig:red_blue} normalized with the quadrupole and octupole power-ratios of the total images (Fig.~\ref{fig:images_tot}) as defined in Eq.~\ref{eq:Pshifted}.
 
\begin{figure}
\centering
\includegraphics[width=9cm]
{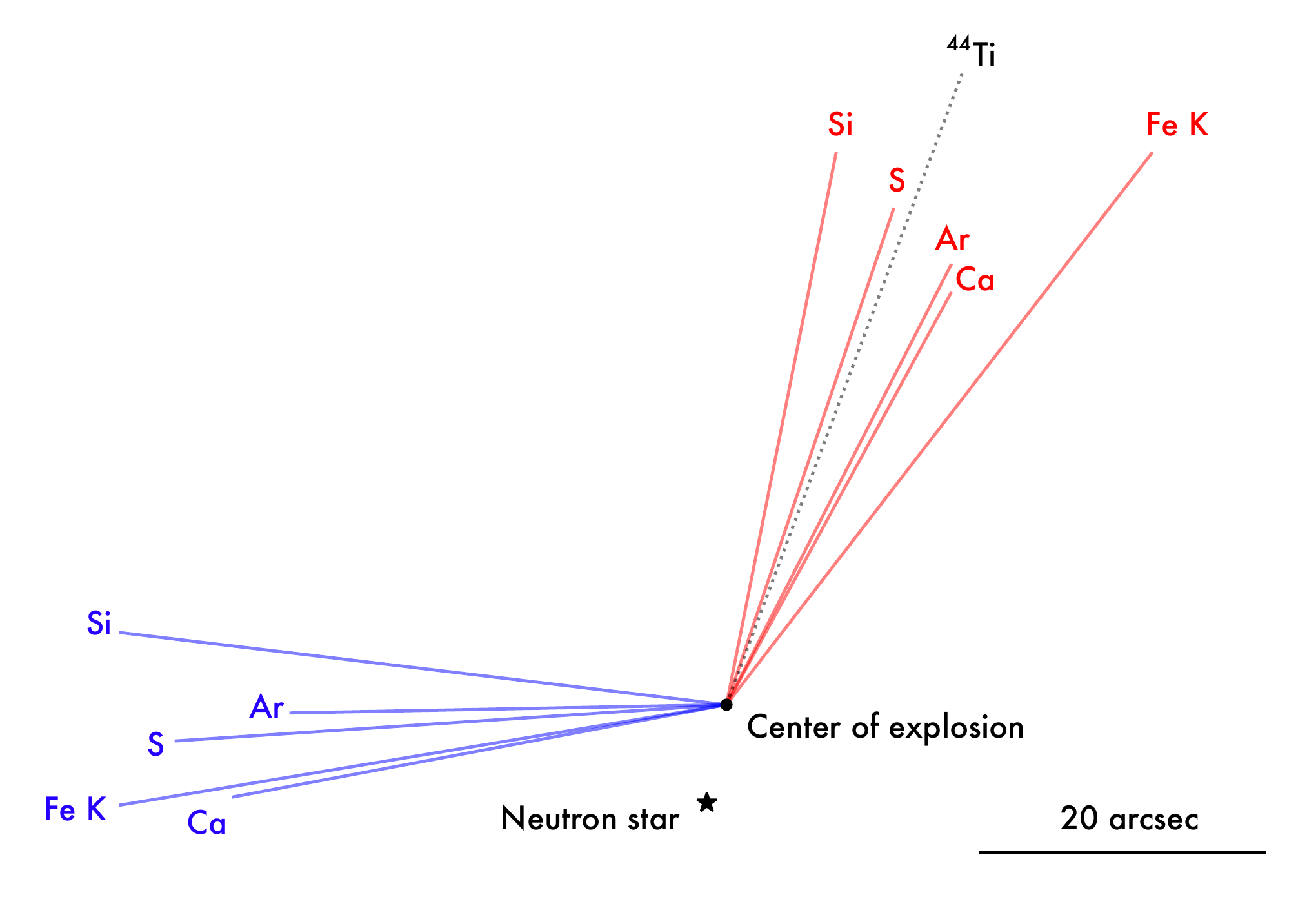}
\caption{Centroids of the blue- and red-shifted parts of each line emission and their distance from the center of explosion of Cas~A. For reference, we added the direction of motion of the $^{44}$Ti in black, as shown in Fig.~13 of \cite{grefenstette17}. Only the direction is relevant as the norm of this specific vector is arbitrary.}
\label{fig:centroids}
\end{figure}

%\begin{figure}
%\centering
%\includegraphics[width=9cm]
%{images/angles.pdf}
%\caption{Angles between the directions of the red- %and blue-shifted centers of emission toward the %center of explosion for each element. The grey %bars are errorbars obtained thanks to the %constrained bootstrap resamplings.}
%\label{fig:angles}
%\end{figure}

\subsection{Discussion on the retrieved spectra}
\label{sub:spectra}

As stated before, it is the spectra retrieved together with the aforementioned images that allow us to identify them as "blue-" or "red-shifted" components. Here we will expand on our reasons for supporting these assertions.  

The spectra in Fig.~\ref{fig:red_blue_images} are superimposed with the theoretical positions of the main emission lines in the energy range. In the case of Si, the retrieved features are shifted to lower or higher energy with respect to the rest-energy positions of the Si~{\sc xiii} and Si~{\sc xiv} lines. Appendix~\ref{ap:velocity} shows that this shifting is not primarily due to an ionization effect as the ratio Si~{\sc xiii}/Si~{\sc xiv} is roughly equal in both cases. The same goes for S, where two lines corresponding to S~{\sc xv} and S~{\sc xvi} are shifted together while keeping a similar ratio. 

%The Ar, Ca and Fe-K cases are single lines shifted around the theoretical position of the line emission, making the "blue-" and "red-shifted" identification straightforward. Fig.~\ref{fig:red_blue_images} also shows that the Si emission is dominated by Si~{\sc xiii}, the S by S~{\sc xv}, the Ar by Ar XVII (Ar XVII has a much higher intensity than S XVII at $1.5$ kev), the Ca by Ca XIX and the Fe-K by Fe XXV (Fe XXV has a much higher intensity than Fe XXIV).

A word on the Ca blue-shifted emission: This component is very weak and in a region where there is a lot of spatial overlap, making it difficult for the algorithm to retrieve. For that reason, the retrieved spectrum has a poorer quality than the others, and it was imperfectly found on some of our constrained bootstrap resamplings. Consequently, we were compelled to run the algorithm on more than a hundred resamplings and to select the accurate ones to obtain a significant envelop around the spectrum obtained on the original data.

\begin{table}
\begin{tabular}{l|l c c c c c}
\hline
\hline
Line & E$_{\rm rest}$ & E$_{\rm red}$ & E$_{\rm blue}$ & $\Delta$V & V$_{\rm red}$ & V$_{\rm blue}$ \\
 & keV & keV & keV & km/s & km/s & km/s \\
 \hline
Si {\sc xiii} & 1.8650 & 1.860 & 1.896 & 5787 & 804 & 4983 \\

Si {\sc xiii}$^{*}$ & 1.8730 & 1.860 & 1.896 & 5762 & 2081 & 3681 \\

S {\sc xv} & 2.4606 & 2.439 & 2.489 & 6092 & 2632 & 3460 \\

Ar {\sc xvii} & 3.1396 & 3.110 & 3.180 & 6684 & 2826 & 3858 \\

Ca {\sc xix} & 3.9024 & 3.880 & 3.967 & 6684 & 1721 & 4963 \\

Fe complex & 6.6605 & 6.599 & 6.726 & 5716 & 2768 & 2948 \\
\hline
\end{tabular}
\caption{Spectral fitting on individual lines and resulting velocities. Si and Fe line rest energy are taken from \citet{delaney10}. The Si XIII$^{*}$ uses a different rest energy, the one needed to match the ACIS and HETG Si velocities discussed in \citet{delaney10}, to illustrate possible ACIS calibration issues. }
\label{tab:linefit}
\end{table}

\subsection{Spectral analysis}
\label{sub:xspec}
%\fabio{section updated }

Using the spectral components retrieved for each data subset shown in Fig.~\ref{fig:red_blue_images}, we carried out a spectral fitting assuming a residual continuum plus line emission in \textit{XSPEC} (power-law + Gauss model).  In this analysis, the errors for each spectral data point are derived from the constraint bootstrap method presented in Appendix \ref{ap:boot}. This constrained bootstrap eliminates a bias introduced by classical bootstrap methods and that is critical to pGMCA, but underestimates the true statistical error. Therefore, no statistical errors on the line centroids are listed in  Table~\ref{tab:linefit} as, in addition, systematic errors associated with ACIS energy calibration are likely to be the dominant source of uncertainty.

The resulting line centroid and equivalent velocity shifts are shown in Table~\ref{tab:linefit}. To transform the shift in energy into a velocity shift, a rest energy is needed. The ACIS CCD spectral resolution does not resolve the line complex and cannot easily disentangle velocity and ionization effects. However, given the range of ionization state observed in Cas~A \citep[with ionization ages of $tau \sim10^{11}-10^{12}$ cm$
^{-3}$ s, see Fig. 2 of][]{hwang12},
there is little effect of ionization on the dominant line for  Si, S, Ar, and Ca, as discussed in more details in Appendix \ref{ap:velocity}. The line rest energy was chosen as the brightest line for a non-equilibrium ionization plasma with a temperature of 1.5 keV temperature and $log(\tau) = 11.3$ cm$^{-3}$ s, the mean values from Fig. 2 of \citet{hwang12}.

For the specific case of the Si {\sc xiii} line, a very large asymmetry in the red/blue-shifted velocities is observed. This could be due to possible energy calibration issues near the Si line as shown by \citet{delaney10} in a comparison of ACIS and HETG line centroid, resulting in a systematic blue-shift effect in ACIS data.
The Si {\sc xiii}$^{*}$ line in Table \ref{tab:linefit} uses a corrected rest line energy to illustrate systematic uncertainties associated with calibration issues.

For the Fe-K complex of lines, we rely on the analysis of \citet{delaney10} who derived an average rest line energy of 6.6605~keV (1.8615 {\AA}) by fitting a spherical expansion model to their three-dimensional ejecta model. We can note that with this spectral analysis, what we measure here is the radial velocity that is flux weighted over the entire image of the associated component. Therefore, we are not probing the velocity at small angular scale but the bulk velocity of the entire component.

With the caveats listed above, we notice an asymmetry in the velocities where ejecta seem to have a higher velocity toward us (blue-shifted) than away from us, even in the case of Si {\sc xiii} after calibration corrections.
A comparison of those results with previous studies and possibles biases are discussed in Sect.~\ref{sect:XX}. The large uncertainties associated with the energy calibration and the choice of rest energy has little impact on the delta between the red- and blue shifted centroids and hence on the $\Delta$V.
We note that all elements show a consistent $\Delta$V of $\sim$6000 km s$^{-1}$.

%Blue component has smaller, compact knots less prone to the averaging effects.

\section{Physical interpretation}
\label{sect:physics}
\subsection{Quantification of ejecta asymmetries}
\label{sub:int_prm}

Fig.~\ref{fig:asym} shows that the distribution of heavier elements is generally more elliptical and more mirror asymmetric than that of lighter elements in Cas~A: O, Si, S, Ar, Ca, and Fe emission all exhibit successively higher levels of both measures of asymmetry. This result is consistent with the recent observational study of Cas~A by \citet{hollandashford20}, suggesting that the pGMCA method accurately extracts information from X-ray data cubes without the complicated and time-consuming step of extracting spectra from hundreds or thousands of small regions and analyzing them individually.

Similar to the results of \citet{hollandashford20} and \citet{hwang12}, Mg emission does not follow the exact same trend as the other elements~: it has roughly an order of magnitude lower elliptical asymmetry ($P_{2}/P_{0}$) than the other elements. In contrast to \citet{hollandashford20} and \citet{hwang12}, our Mg image (as shown in Fig.~\ref{fig:other_images}) presents a morphology highly different from that of the Fe L; we believe that the pGMCA was able to retrieve the Mg spatial distribution with little continuum or Fe contamination.  %We agree with \cite{hollandashford20} that the Mg emission is likely contaminated by other nearby lines: e.g., Fig.~\ref{fig:red_blue_images} shows the 1.41 keV FeXXIII emission line is near the center of what we identified as blue-shifted Mg emission (FIXME NOT ANYMORE). Furthermore, at these lower energies, ionization timescale has a significant effects on which lines dominate the spectra. As shown in Fig.~\ref{fig:xspec_model}, zero-velocity Mg emission can be centered around 1.33 or 1.48 keV depending on the ionization timescale. Thus, what we identify as blue-shifted Mg (FIXME NOT ANYMORE) at $n_eT<$ 10$^{11}$ cm$^{-3}$ s may in fact be red-shifted Mg emission at $n_eT>$ 10$^{11}$ cm$^{-3}$ s. 
%\indent We offer a similar explanation/caution for the asymmetry result of the Fe-L distribution: it is impossible to disentangle the various Fe XXII and Fe XXIV lines present in the 1-1.2 keV range. Thus, what the pGMCA labels as blue-shifted Fe-L (FIXME NOT ANYMORE) emission is likely contaminated by red-shifted Fe-L (FIXME NOT ANYMORE) emission from, e.g., the 1.11, 1.13, and 1.17 keV Fe lines. In addition, for the Fe-L emission in particular, it is difficult to disentangle velocity-shifted ejecta from ejecta at different ionization timescales. Fig.~\ref{fig:xspec_model} shows that the emission we label as red-shifted matches a sample 2keV spectra with an ionization timescale of 10$^{11}$ cm$^{-3}$ s, and the blue-shifted spectra matches a sample 2keV spectra with a higher ionization timescale of 10$^{12}$ cm$^{-3}$ s.

Fig.~\ref{fig:red_blue} presents the relative ellipticity/elongation and mirror asymmetries of the blue- and red-shifted ejecta emission compared to the total ejecta images (Fig.~\ref{fig:images_tot}). A value of ``1'' indicates that the velocity-shifted ejecta has equivalent levels of asymmetry as the full bandpass emission. In the cases where we can clearly disentangle the red- and blue-shifted emission (i.e.Si, S, Ar, Ca, and Fe-K, described in previous paragraphs), we see that the red-shifted ejecta emission is less asymmetric than the blue-shifted emission. This holds true both for elliptical asymmetry $P_{2}/P_{0}$ and mirror asymmetry $P_{3}/P_{0}$. Thus, we could physically describe the red-shifted ejecta distribution as a broad, relatively symmetric plume, whereas the blue-shifted ejecta is concentrated into dense knots. This interpretation matches with the observation that most of the X-ray emission is from the red-shifted ejecta, as we can also see in the flux ratios shown in Table~\ref{tab:frac} and in the images of Fig.~\ref{fig:red_blue_images}, suggesting that there was more mass ejected away from the observer, NS, and blue-shifted ejecta knot. We note that there is not a direct correlation between ejecta mass and X-ray emission due to the position of the reverse shock, the plasma temperature and ionization timescale, but the indication that most of the X-ray emission is red-shifted is consistent with our knowledge of the $^{44}$Ti distribution (see Sect. \ref{sect:ti} for a more detailed discussion).

Furthermore, in all cases, the red-shifted ejecta emission is more circularly symmetric than the total images, and the blue-shifted ejecta is more elliptical and elongated than the total images. Moreover, the red-shifted ejecta is more mirror symmetric than the blue-shifted ejecta, though both the red-shifted and blue-shifted Si are more mirror asymmetric than the total image. The latter result may suggest that the red-shifted and blue-shifted Si images' asymmetries sum together such that the total Si image appears more mirror symmetric than the actual distribution of the Si.

\subsection{Three-dimensional distribution of heavy elements}
\label{sub:int_velocity}

%\adrien{Fig. \ref{fig:centroids} shows the centroids of the blue- and red-shifted parts of each line emission and their distance relative to the center of explosion of Cas~A, hence giving information about their mean direction and bulk motion. The blue- and red-shifted parts of a given element are obviously not diametrically opposed, proving wrong the idea of a jet/counter-jet explosion mechanism. Table~\ref{tab:frac} supports this idea by showing the asymmetries between blue- and red-shifted parts, the latter representing a far more important fraction of the total flux for all line emissions. We will see in Section \ref{sect:ti} that this is consistent with previous works on Cas~A investigating the $^{44}$Ti distribution with NuSTAR data.}

Fig.~\ref{fig:centroids} shows the centroids of the blue- and red-shifted parts of each emission line relative to the center-of-explosion of Cas~A, revealing the bulk three-dimensional distribution of each component. We note that this figure was only made using the centroids of the red- and blue-shifted images retrieved by pGMCA, without using the PRM method. We can see the red-shifted ejecta is mainly moving in a similar direction (toward the northwest), while all the blue-shifted ejecta is moving toward the east. As discussed in Section~\ref{sect:ti}, this result is consistent with previous works on Cas~A investigating the $^{44}$Ti distribution with {\it NuSTAR} data \citep{grefenstette17}.

The blue-shifted ejecta is clearly moving in a different direction than the red-shifted ejecta, but not directly opposite to it. The angles between the blue- and red-shifted components are all between $90^\circ$ and $140^\circ$. This finding provides evidence against a jet and counter-jet explosion mechanism being responsible for the explosion and resulting in the expansion of ejecta in Cas~A (e.g., \citealt{fesen01,hines04,schure08}). However, this could support the idea of a jittering jets mechanism, wherein multiple jets can be launched in different directions, as described in \cite{Bear_2018}. In the same paper, it is argued that jets clear out ejecta material, so a NE-SW jet would be consistent with our identified ejecta motions, as it would be perpendicular to our redshifted ejecta and still at a ~45 degree angle from our blueshifted ejecta.

We can also note a trend where heavier elements exhibit increasingly larger opening angles than lighter elements, from Si showing a $90^\circ$ angle to Ca and Fe that show opening angles of about $130-140^\circ$.%, which can give insights on asymmetry generation in the core of the SN close to the proto-NS. This is consistent with recent simulations (e.g., \citealt{wongwathanarat13,wongwathanarat17,janka17}) which predict that asymmetric explosion processes result in the heaviest ejecta synthesized closest to the core exhibiting the strongest levels of asymmetry.}

\subsection{Velocities of red- and blue-shifted structures}
\label{sect:XX}

By fitting the line centroids, we obtained the velocities discussed in Sect. \ref{sub:xspec}. As stated before, the effects of ionization on possible "imposter velocities" are discussed in Appendix~\ref{ap:velocity}. Our derived velocities showed higher values for the blue-shifted component than for the red-shifted one for all elements. Those results are in disagreement with spectroscopic studies and in agreement with some others. On the one hand, the X-ray studies of individual regions \citet{willingale02} (Fig. 8, XMM-Newton EPIC cameras) and \citet{delaney10} (Fig. 10 and 11, Chandra ACIS and HETG instruments) 
indicate higher velocities for the red-shifted component.
But on the other hand, the highest velocity measured in the $^{44}$Ti NuSTAR analysis is for the blue-shifted component \citep[Table 3 of ][]{grefenstette17}. 
We note that the comparison is not straightforward as the methods being used are different. Our method measures a flux weighted average velocity for each well separated component whereas in the X-ray studies previously mentioned, a single gaussian model is fitted to the spectrum extracted in each small-scale region. In regions where both red- and blue-shifted ejecta co-exist (see Fig. \ref{fig:red_blue_images}), the Gaussian fit will provide a flux weighted average velocity value of the two components as they are not resolved with ACIS. 
As the red-shifted component is brighter in average, a systematic bias that would reduce the blue velocities could exist.
This could be the case in the southeastern region where most of the blue-shifted emission is observed and where a significant level of red-shifted emission is also seen.
Besides this, calibration issues may also play an important role.
Although the GMCA method was successful in retrieving the centroid energy of nearby emission lines using a simple toy model \citep[Fig.~7 in ]{picquenot:hal-02160434}, we do not rule out that the higher velocity of the blue-shifted component is an artifact of the method. Further tests of the method with the help of synthetic X-ray observations using numerical simulations could shed light on this issue.

\subsection{Neutron star kick direction}
\label{sub:int_NSvelocity}
%The NS in Cas~A is located SE of the explosion site, moving at a velocity of $\sim$340 km s$^{-1}$ SE in the plane of the sky \citep{thorstensen01}. This is roughly opposite to the direction of bulk ejecta motion \citep{hollandashford17,katsuda18}, and almost exactly opposite to the bulk of heavy element (Ar, Ca, Ti, Fe) motion \citep{grefenstette17,hollandashford20}. This correlation provides evidence for the theory that NSs are kicked opposite to the direction of bulk ejecta motion consistent with conservation of momentum with the ejecta \citep{wongwathanarat13,muller16,bruenn16, janka17}. Specifically, observations have provided evidence for the `Gravitational Tugboat Mechanism' of generating NS kicks asymmetries proposed by \cite{wongwathanarat13, janka17}, where the NS is gravitationally accelerated by the slower moving ejecta clumps, opposite to the bulk ejecta motion.

The NS in Cas~A is located southeast of the explosion site, moving at a velocity of $\sim$340 km s$^{-1}$ southeast in the plane of the sky \citep{thorstensen01,Fesen_2006}. In \citet{hwang12} it was stated that, contrary to expectations, the Fe structure was not observed to recoil in the opposite direction to the NS. Here, thanks to our ability to disentangle red- and blue-shifted structures, we find that the red-shifted ejecta is moving nearly opposite the NS : The angles between the red-shifted structures and the NS tangential motion range between $154^\circ$ (Fe) and $180^\circ$ (Si). Table.~\ref{tab:frac} also shows that the bulk emission is from red-shifted ejecta \citep[consistent with][]{2013ApJ...772..134M}. This correlation is consistent with theoretical predictions that NSs are kicked opposite to the direction of bulk ejecta motion, in adequation with conservation of momentum with the ejecta \citep{PhysRevLett.76.352,Nordhaus_2010,Nordhaus_2012,wongwathanarat13,muller16,bruenn16,janka17}. \cite{wongwathanarat13, janka17} also proposed the existence of a "gravitational tugboat mechanism" where the NS is gravitationally accelerated by the slower moving ejecta clumps, opposite to the bulk ejecta motion. However, in the simulations of \cite{Nordhaus_2012}, the direction of the NS is mainly due to recoil, and the effects of gravitation are negligible. The asymmetric neutrino emissions can also play a role in generating a NS kick opposite to the bulk ejecta motion \citep{2019ApJ...880L..28N,2019MNRAS.488L.114F,2021MNRAS.502.2319F}. However, the Zn distribution, that we could not observe in this study, would be a better marker to test this last hypothesis.

It is impossible to calculate the NS line-of-sight motion by examining the NS alone as its spectra contains no lines to be Doppler-shifted. However, limits on its three-dimensional motion can be placed by assuming it moves opposite the bulk of ejecta and examining the bulk three-dimensional motion of ejecta. \cite{grefenstette17} studied Ti emission in Cas~A and found that the bulk Ti emission was tilted 58$^\circ$ into the plane of the sky away from the observer, implying that the NS is moving 58$^\circ$ out of the plane of the sky toward the observer. This finding is supported by three-dimensional simulations of a Type IIb progenitor by \cite{wongwathanarat17} and \cite{jerkstrand20}, which suggested that the NS is moving out of the plane of the sky with an angle of $\sim$ 30$^\circ$. 

The results of this paper support the hypothesis that, if the NS is moving away from the bulk of ejecta motion, the NS is moving toward us. Furthermore, we could tentatively conclude that the NS was accelerated toward the more slowly moving blue-shifted ejecta, which would further support the gravitational tugboat mechanism or a kick only caused by momentum conservation. The strong levels of asymmetry exhibited by the blue-shifted emission combined with the lower flux would imply that the blue-shifted ejecta is split into  relatively small ejecta clumps, one of which would possibly be the source of the neutron star's gravitational acceleration. However, the velocities determined in Table~\ref{tab:linefit} contradict this hypothesis as the blue-shifted clumps seem to move faster. 

\begin{figure}
\centering
\includegraphics[width=8cm]
{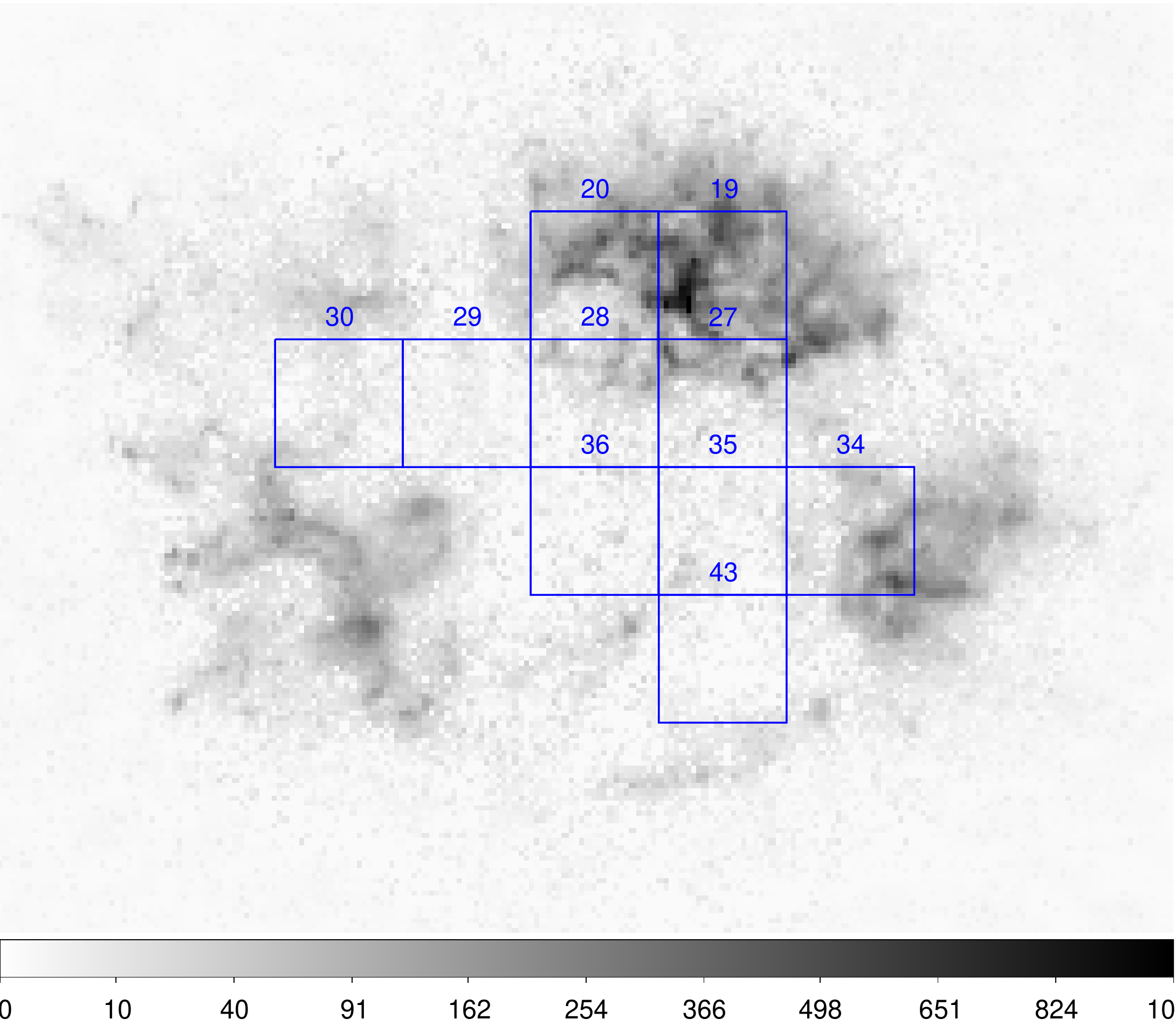}
\caption{Counts image of the Fe-K red-shifted component overlaid with the extraction regions used for the $^{44}$Ti NuSTAR study of \citet{grefenstette17}. The regions 19 and 20, which dominate our image in terms of flux, have respective velocities moving away from the observer of 2300 $\pm$ 1400 and 3200 $\pm$ 500 km sec$^{-1}$.}
\label{fig:nustar}
\end{figure}

%\fabio{idea here is that if we see 44Ti and Iron in the same regions, they could have similar velocities. So region 19 which is dominating our X-ray images might have a velocity of $\sim$ 2000 km/s hence a redshift of the Fe-K of about 40 eV. Also no blue-shifted Iron is detected in the SE in NuSTAR. }

\subsection{Comparison with $^{44}$Ti }
\label{sect:ti}

\begin{table}
\centering
\begin{tabular}{c|c c}
  \hline
  \hline
    & Red-shifted part & Blue-shifted part \\
  \hline
  Si & 0.60 & 0.40 \\
  
  S & 0.61 & 0.39 \\
  
  Ar & 0.63 & 0.37 \\
  
  Ca & 0.80 & 0.20 \\
  
  Fe-K & 0.70 & 0.30 \\
  \hline
\end{tabular}
\caption{Fractions of the counts in the total image that belong to the red-shifted or the blue-shifted parts, for each line.}
\label{tab:frac}

\end{table}

$^{44}$Ti is a product of Si burning and is thought to be synthesized in close proximity with iron.
The $^{44}$Ti spatial distribution has been studied via its radioactive decay with the {\it NuSTAR} telescope and revealed that most of the material is red-shifted and does not seem to follow the Fe-K X-ray emission \citep{grefenstette14,grefenstette17}.
In our study, we have found that 70\%  of the Fe-K X-ray emission (see Table \ref{tab:frac}) is red-shifted and that the mean direction of the Fe-K red-shifted emission shown in Fig.~\ref{fig:centroids} is compatible with that of the $^{44}$Ti as determined in Fig.~13 of \cite{grefenstette17}. 
Yet, we can see the mean $^{44}$Ti direction is not perfectly aligned with the mean red-shifted Fe-K direction. This may be caused by the fact that the Fe-K emission is tracing only the reverse shock-heated material and may not reflect the true distribution of Fe, whereas $^{44}$Ti emission is from radioactive decay and thus reflects the true distribution of Ti.

In Fig.~\ref{fig:nustar}, we overlay the ten regions where  \citet{grefenstette17} detected $^{44}$Ti with our red-shifted component image. 
The regions 19 and 20 (which dominate our Fe-K red-shifted component image)  have respective $^{44}$Ti velocities of 2300 $\pm$ 1400 and 3200 $\pm$ 500 km s$^{-1}$, values that are compatible with our measured value of $\sim$2800 km s$^{-1}$ shown in Table \ref{tab:linefit}.

Concerning our Fe-K blue-shifted component map, its X-ray emission is fainter and located mostly in the southeast of the source (see Fig~\ref{fig:red_blue_images}). This southeastern X-ray emission is spatially coincident with region 46 in the Fig.~2 {\it NuSTAR} map of \citet{grefenstette17}, not plotted in our Fig~\ref{fig:nustar} as the $^{44}$Ti emission was found to be below the detection threshold.

We note that blue-shifted $^{44}$Ti emission is harder to detect for {\it NuSTAR} than a red-shifted one as it is intrinsically fainter. In addition, any blue-shifted emission of the 78.32~keV $^{44}$Ti line places it outside the {\it NuSTAR} bandpass, precluding detection of one of the two radioactive decay lines in this case.

\section{Conclusions}

By using a new methodology and applying it to Cas~A {\it Chandra} X-ray data, we were able to revisit the mapping of the heavy elements and separate them into a red- and a blue-shifted parts, allowing us to investigate the three-dimensional morphology of the SNR. These new maps and the associated spectra could then be used to quantify the asymmetries of each component, their mean direction and their velocity. The main findings of the paper are consistent with the general results found in the previous studies cited in Part \ref{sect:physics}, and are summarized below:
\begin{itemize}
    \item \textbf{Morphological asymmetries:} An extensive study of the asymmetries shows the distribution of heavier elements is generally more elliptical and mirror asymmetric in Cas~A, which is consistent with simulation predictions. For the elements we were able to separate into a red- and a blue-shifted parts (Si, S, Ar, Ca, Fe), it appears that the red-shifted ejecta is less asymmetric than the blue-shifted one. The red-shifted ejecta can then be described as a broad, relatively symmetric plume, while the blue-shifted ejecta can be seen as concentrated into dense knots. Most of the emission from each element is red-shifted, implying there was more mass ejected away from the observer, which agrees with past studies.
    \item \textbf{Three-dimensional distribution:} The mean directions of the red- and blue- shifted parts of each element are clearly not diametrically opposed, disfavouring the idea of a jet/counter-jet explosion mechanism, but not that of a jittering jets mechanism.% The angles between the red- and blue-shifted parts become wider with increasing element mass, indicating that elements formed closer to the core and proto-NS experience stronger asymmetric forces.
    \item \textbf{NS velocity:} We find that the NS is moving most opposite to the direction of the red-shifted ejecta that forms the bulk of the ejecta emission. This supports the idea of a recoil mechanism generating NS kicks through a process consistent with conservation of momentum between NS and ejecta, possibly including gravitational acceleration from the slower moving clumps ("Gravitational Tug Boat Mechanism"), or the effects of asymmetric neutrino emissions. This result implies that the NS is moving toward us, which is consistent with the findings of past studies. However, we find the blue-shifted clumps to be faster than the red-shifted ones, which is not consistent with the gravitational tug-boat mechanism's prediction that the NS is moving opposite to the faster ejecta.
    \item \textbf{Comparison with $^{44}$Ti:} Our finding that the bulk of ejecta is red-shifted and moving NW is consistent with the $^{44}$Ti distribution from NuSTAR observations. Its direction is similar to that of the red-shifted Fe-K emission, but a slight difference could be explained by the fact that the Fe-K only traces the reverse shock-heated ejecta and not the full distribution of the Fe ejecta.
    
\end{itemize}

The component separation method presented here enabled a three-dimensional view of the Cas~A ejecta despite the low energy resolution of the {\it Chandra} CCDs by separating entangled components all at once, without the need of a detailed spectral analysis on hundreds of regions.  In the future, X-ray microcalorimeters will enable kinematic measurements of X-ray emitting ejecta in many more SNRs. In its short operations, the {\it Hitomi} mission demonstrated these powerful capabilities. In particular, in a brief 3.7-ks observation, it revealed that the SNR N132D had highly red-shifted Fe emission with a velocity of $\sim$800~km~s$^{-1}$ without any blue-shifted component, suggesting the Fe-rich ejecta was ejected asymmetrically \citep{hitomi18}. The upcoming replacement X-ray Imaging and Spectroscopy Mission {\it XRISM} will offer 5--7~eV energy resolution with 30\arcsec\ pixels over a 3\arcmin\ field of view \citep{tashiro18}. In the longer term, {\it Athena} and {\it Lynx} will combine this superb spectral resolution with high angular resolution, fostering a detailed, three-dimensional view of SNRs that will revolutionize our understanding of explosions \citep{lopez19,williams19}. While the new instruments will provide a giant leap forward in terms of data quality, development of new analysis methods are needed in order to maximize the scientific return of  next generation telescopes.

\begin{acknowledgements}
This research made use of Astropy,\footnote{http://www.astropy.org} a community-developed core Python package for Astronomy \citep{astropy:2013, astropy:2018} and of gammapy,\footnote{https://www.gammapy.org} a community-developed core Python package for TeV gamma-ray astronomy \citep{gammapy:2017, gammapy:2019}.
We also acknowledge the use of Numpy \citep{oliphant2006guide} and Matplotlib \citep{Hunter:2007}.
Thanks also to Adam Burrows and Hiroki Nagakura for their precious help during the redaction of the revised version.

\end{acknowledgements}

   \bibliographystyle{aa} % style aa.bst
   \bibliography{asymmetries_casA} % your references Yourfile.bib

\begin{thebibliography}{63}
\expandafter\ifx\csname natexlab\endcsname\relax\def\natexlab#1{#1}\fi

\bibitem[{{Alarie} {et~al.}(2014){Alarie}, {Bilodeau}, \&
  {Drissen}}]{2014MNRAS.441.2996A}
{Alarie}, A., {Bilodeau}, A., \& {Drissen}, L. 2014, \mnras, 441, 2996

\bibitem[{{Astropy Collaboration} {et~al.}(2013){Astropy Collaboration},
  {Robitaille}, {Tollerud}, {Greenfield}, {Droettboom}, {Bray}, {Aldcroft},
  {Davis}, {Ginsburg}, {Price-Whelan}, {Kerzendorf}, {Conley}, {Crighton},
  {Barbary}, {Muna}, {Ferguson}, {Grollier}, {Parikh}, {Nair}, {Unther},
  {Deil}, {Woillez}, {Conseil}, {Kramer}, {Turner}, {Singer}, {Fox}, {Weaver},
  {Zabalza}, {Edwards}, {Azalee Bostroem}, {Burke}, {Casey}, {Crawford},
  {Dencheva}, {Ely}, {Jenness}, {Labrie}, {Lim}, {Pierfederici}, {Pontzen},
  {Ptak}, {Refsdal}, {Servillat}, \& {Streicher}}]{astropy:2013}
{Astropy Collaboration}, {Robitaille}, T.~P., {Tollerud}, E.~J., {et~al.} 2013,
  \aap, 558, A33

\bibitem[{Bear \& Soker(2018)}]{Bear_2018}
Bear, E. \& Soker, N. 2018, Monthly Notices of the Royal Astronomical Society

\bibitem[{{Bobin} {et~al.}(2020){Bobin}, {Hamzaoui}, {Picquenot}, \&
  {Acero}}]{9215040}
{Bobin}, J., {Hamzaoui}, I.~E., {Picquenot}, A., \& {Acero}, F. 2020, IEEE
  Transactions on Image Processing, 29, 9429

\bibitem[{{Bobin} {et~al.}(2015){Bobin}, {Rapin}, {Larue}, \&
  {Starck}}]{bobin15}
{Bobin}, J., {Rapin}, J., {Larue}, A., \& {Starck}, J.-L. 2015, IEEE
  Transactions on Signal Processing, 63, 1199

\bibitem[{{Bruenn} {et~al.}(2016){Bruenn}, {Lentz}, {Hix}, {Mezzacappa},
  {Harris}, {Messer}, {Endeve}, {Blondin}, {Chertkow}, {Lingerfelt},
  {Marronetti}, \& {Yakunin}}]{bruenn16}
{Bruenn}, S.~W., {Lentz}, E.~J., {Hix}, W.~R., {et~al.} 2016, \apj, 818, 123

\bibitem[{{Buote} \& {Tsai}(1995)}]{1995ApJ...452..522B}
{Buote}, D.~A. \& {Tsai}, J.~C. 1995, \apj, 452, 522

\bibitem[{Burrows \& Hayes(1996)}]{PhysRevLett.76.352}
Burrows, A. \& Hayes, J. 1996, Phys. Rev. Lett., 76, 352

\bibitem[{Burrows {et~al.}(2019)Burrows, Radice, Vartanyan, Nagakura, Skinner,
  \& Dolence}]{Burrows_2019}
Burrows, A., Radice, D., Vartanyan, D., {et~al.} 2019, Monthly Notices of the
  Royal Astronomical Society, 491, 2715–2735

\bibitem[{Burrows \& Vartanyan(2021)}]{Burrows_Vartanyan20}
Burrows, A. \& Vartanyan, D. 2021, Nature, 589, 29

\bibitem[{{Deil} {et~al.}(2017){Deil}, {Zanin}, {Lefaucheur}, {Boisson},
  {Khelifi}, {Terrier}, {Wood}, {Mohrmann}, {Chakraborty}, {Watson},
  {Lopez-Coto}, {Klepser}, {Cerruti}, {Lenain}, {Acero}, {Djannati-Ata{\"\i}},
  {Pita}, {Bosnjak}, {Trichard}, {Vuillaume}, {Donath}, {Consortium}, {King},
  {Jouvin}, {Owen}, {Sipocz}, {Lennarz}, {Voruganti}, {Spir-Jacob}, {Ruiz}, \&
  {Arribas}}]{gammapy:2017}
{Deil}, C., {Zanin}, R., {Lefaucheur}, J., {et~al.} 2017, in International
  Cosmic Ray Conference, Vol. 301, 35th International Cosmic Ray Conference
  (ICRC2017), 766

\bibitem[{{DeLaney} {et~al.}(2010){DeLaney}, {Rudnick}, {Stage}, {Smith},
  {Isensee}, {Rho}, {Allen}, {Gomez}, {Kozasa}, {Reach}, {Davis}, \&
  {Houck}}]{delaney10}
{DeLaney}, T., {Rudnick}, L., {Stage}, M.~D., {et~al.} 2010, \apj, 725, 2038

\bibitem[{DeLaney \& Satterfield(2013)}]{delaney13}
DeLaney, T. \& Satterfield, J. 2013

\bibitem[{{Dewey}(2010)}]{dewey2010}
{Dewey}, D. 2010, \ssr, 157, 229

\bibitem[{Efron(1979)}]{befron79}
Efron, B. 1979, Volume 7, Number 1, 1-26

\bibitem[{Ferrand {et~al.}(2019)Ferrand, Warren, Ono, Nagataki, Röpke, \&
  Seitenzahl}]{Ferrand_2019}
Ferrand, G., Warren, D.~C., Ono, M., {et~al.} 2019, The Astrophysical Journal,
  877, 136

\bibitem[{{Fesen}(2001)}]{fesen01}
{Fesen}, R.~A. 2001, \apjs, 133, 161

\bibitem[{Fesen {et~al.}(2006)Fesen, Hammell, Morse, Chevalier, Borkowski,
  Dopita, Gerardy, Lawrence, Raymond, \& van~den Bergh}]{Fesen_2006}
Fesen, R.~A., Hammell, M.~C., Morse, J., {et~al.} 2006, The Astrophysical
  Journal, 645, 283

\bibitem[{{Fesen} {et~al.}(2001){Fesen}, {Morse}, {Chevalier}, {Borkowski},
  {Gerardy}, {Lawrence}, \& {van den Bergh}}]{2001AJ....122.2644F}
{Fesen}, R.~A., {Morse}, J.~A., {Chevalier}, R.~A., {et~al.} 2001, \aj, 122,
  2644

\bibitem[{{Foster} {et~al.}(2012){Foster}, {Ji}, {Smith}, \&
  {Brickhouse}}]{2012ApJ...756..128F}
{Foster}, A.~R., {Ji}, L., {Smith}, R.~K., \& {Brickhouse}, N.~S. 2012, \apj,
  756, 128

\bibitem[{{Fujimoto} \& {Nagakura}(2019)}]{2019MNRAS.488L.114F}
{Fujimoto}, S.-i. \& {Nagakura}, H. 2019, \mnras, 488, L114

\bibitem[{{Fujimoto} \& {Nagakura}(2021)}]{2021MNRAS.502.2319F}
{Fujimoto}, S.-i. \& {Nagakura}, H. 2021, \mnras, 502, 2319

\bibitem[{Gessner \& Janka(2018)}]{Gessner_2018}
Gessner, A. \& Janka, H.-T. 2018, The Astrophysical Journal, 865, 61

\bibitem[{{Gotthelf} {et~al.}(2001){Gotthelf}, {Koralesky}, {Rudnick}, {Jones},
  {Hwang}, \& {Petre}}]{Gotthelf01}
{Gotthelf}, E.~V., {Koralesky}, B., {Rudnick}, L., {et~al.} 2001, \apjl, 552,
  L39

\bibitem[{{Greco, Emanuele} {et~al.}(2020){Greco, Emanuele}, {Vink, Jacco},
  {Miceli, Marco}, {Orlando, Salvatore}, {Domcek, Vladimir}, {Zhou, Ping},
  {Bocchino, Fabrizio}, \& {Peres, Giovanni}}]{greco20}
{Greco, Emanuele}, {Vink, Jacco}, {Miceli, Marco}, {et~al.} 2020, A\&A, 638,
  A101

\bibitem[{{Grefenstette} {et~al.}(2017){Grefenstette}, {Fryer}, {Harrison},
  {Boggs}, {DeLaney}, {Laming}, {Reynolds}, {Alexander}, {Barret},
  {Christensen}, {Craig}, {Forster}, {Giommi}, {Hailey}, {Hornstrup},
  {Kitaguchi}, {Koglin}, {Lopez}, {Mao}, {Madsen}, {Miyasaka}, {Mori}, {Perri},
  {Pivovaroff}, {Puccetti}, {Rana}, {Stern}, {Westergaard}, {Wik}, {Zhang}, \&
  {Zoglauer}}]{grefenstette17}
{Grefenstette}, B.~W., {Fryer}, C.~L., {Harrison}, F.~A., {et~al.} 2017, \apj,
  834, 19

\bibitem[{{Grefenstette} {et~al.}(2014){Grefenstette}, {Harrison}, {Boggs},
  {Reynolds}, {Fryer}, {Madsen}, {Wik}, {Zoglauer}, {Ellinger}, {Alexander},
  {An}, {Barret}, {Christensen}, {Craig}, {Forster}, {Giommi}, {Hailey},
  {Hornstrup}, {Kaspi}, {Kitaguchi}, {Koglin}, {Mao}, {Miyasaka}, {Mori},
  {Perri}, {Pivovaroff}, {Puccetti}, {Rana}, {Stern}, {Westergaard}, \&
  {Zhang}}]{grefenstette14}
{Grefenstette}, B.~W., {Harrison}, F.~A., {Boggs}, S.~E., {et~al.} 2014, \nat,
  506, 339

\bibitem[{{Hines} {et~al.}(2004){Hines}, {Rieke}, {Gordon}, {Rho}, {Misselt},
  {Woodward}, {Werner}, {Krause}, {Latter}, {Engelbracht}, {Egami}, {Kelly},
  {Muzerolle}, {Stansberry}, {Su}, {Morrison}, {Young}, {Noriega-Crespo},
  {Padgett}, {Gehrz}, {Polomski}, {Beeman}, \& {Haller}}]{hines04}
{Hines}, D.~C., {Rieke}, G.~H., {Gordon}, K.~D., {et~al.} 2004, \apjs, 154, 290

\bibitem[{{Hitomi Collaboration} {et~al.}(2018){Hitomi Collaboration},
  {Aharonian}, {Akamatsu}, {Akimoto}, {Allen}, {Angelini}, {Audard}, {Awaki},
  {Axelsson}, {Bamba}, {Bautz}, {Blandford}, {Brenneman}, {Brown}, {Bulbul},
  {Cackett}, {Chernyakova}, {Chiao}, {Coppi}, {Costantini}, {de Plaa}, {de
  Vries}, {den Herder}, {Done}, {Dotani}, {Ebisawa}, {Eckart}, {Enoto}, {Ezoe},
  {Fabian}, {Ferrigno}, {Foster}, {Fujimoto}, {Fukazawa}, {Furuzawa},
  {Galeazzi}, {Gallo}, {Gandhi}, {Giustini}, {Goldwurm}, {Gu}, {Guainazzi},
  {Haba}, {Hagino}, {Hamaguchi}, {Harrus}, {Hatsukade}, {Hayashi}, {Hayashi},
  {Hayashida}, {Hiraga}, {Hornschemeier}, {Hoshino}, {Hughes}, {Ichinohe},
  {Iizuka}, {Inoue}, {Inoue}, {Ishida}, {Ishikawa}, {Ishisaki}, {Iwai},
  {Kaastra}, {Kallman}, {Kamae}, {Kataoka}, {Katsuda}, {Kawai}, {Kelley},
  {Kilbourne}, {Kitaguchi}, {Kitamoto}, {Kitayama}, {Kohmura}, {Kokubun},
  {Koyama}, {Koyama}, {Kretschmar}, {Krimm}, {Kubota}, {Kunieda}, {Laurent},
  {Lee}, {Leutenegger}, {Limousin}, {Loewenstein}, {Long}, {Lumb}, {Madejski},
  {Maeda}, {Maier}, {Makishima}, {Markevitch}, {Matsumoto}, {Matsushita},
  {McCammon}, {McNamara}, {Mehdipour}, {Miller}, {Miller}, {Mineshige},
  {Mitsuda}, {Mitsuishi}, {Miyazawa}, {Mizuno}, {Mori}, {Mori}, {Mukai},
  {Murakami}, {Mushotzky}, {Nakagawa}, {Nakajima}, {Nakamori}, {Nakashima},
  {Nakazawa}, {Nobukawa}, {Nobukawa}, {Noda}, {Odaka}, {Ohashi}, {Ohno},
  {Okajima}, {Ota}, {Ozaki}, {Paerels}, {Paltani}, {Petre}, {Pinto}, {Porter},
  {Pottschmidt}, {Reynolds}, {Safi-Harb}, {Saito}, {Sakai}, {Sasaki}, {Sato},
  {Sato}, {Sato}, {Sato}, {Sawada}, {Schartel}, {Serlemtsos}, {Seta},
  {Shidatsu}, {Simionescu}, {Smith}, {Soong}, {Stawarz}, {Sugawara}, {Sugita},
  {Szymkowiak}, {Tajima}, {Takahashi}, {Takahashi}, {Takeda}, {Takei},
  {Tamagawa}, {Tamura}, {Tanaka}, {Tanaka}, {Tanaka}, {Tashiro}, {Tawara},
  {Terada}, {Terashima}, {Tombesi}, {Tomida}, {Tsuboi}, {Tsujimoto}, {Tsunemi},
  {Tsuru}, {Uchida}, {Uchiyama}, {Uchiyama}, {Ueda}, {Ueda}, {Uno}, {Urry},
  {Ursino}, {Watanabe}, {Werner}, {Wilkins}, {Williams}, {Yamada}, {Yamaguchi},
  {Yamaoka}, {Yamasaki}, {Yamauchi}, {Yamauchi}, {Yaqoob}, {Yatsu}, {Yonetoku},
  {Zhuravleva}, \& {Zoghbi}}]{hitomi18}
{Hitomi Collaboration}, {Aharonian}, F., {Akamatsu}, H., {et~al.} 2018, \pasj,
  70, 16

\bibitem[{{Holland-Ashford} {et~al.}(2020){Holland-Ashford}, {Lopez}, \&
  {Auchettl}}]{hollandashford20}
{Holland-Ashford}, T., {Lopez}, L.~A., \& {Auchettl}, K. 2020, \apj, 889, 144

\bibitem[{Hunter(2007)}]{Hunter:2007}
Hunter, J.~D. 2007, Computing in Science \& Engineering, 9, 90

\bibitem[{{Hwang} \& {Laming}(2012)}]{hwang12}
{Hwang}, U. \& {Laming}, J.~M. 2012, \apj, 746, 130

\bibitem[{{Hwang} {et~al.}(2004){Hwang}, {Laming}, {Badenes}, {Berendse},
  {Blondin}, {Cioffi}, {DeLaney}, {Dewey}, {Fesen}, {Flanagan}, {Fryer},
  {Ghavamian}, {Hughes}, {Morse}, {Plucinsky}, {Petre}, {Pohl}, {Rudnick},
  {Sankrit}, {Slane}, {Smith}, {Vink}, \& {Warren}}]{hwang04}
{Hwang}, U., {Laming}, J.~M., {Badenes}, C., {et~al.} 2004, \apjl, 615, L117

\bibitem[{{Janka}(2017)}]{janka17}
{Janka}, H.-T. 2017, \apj, 837, 84

\bibitem[{Janka {et~al.}(2016)Janka, Melson, \& Summa}]{janka16}
Janka, H.-T., Melson, T., \& Summa, A. 2016, Annual Review of Nuclear and
  Particle Science, 66, 341

\bibitem[{{Jerkstrand} {et~al.}(2020){Jerkstrand}, {Wongwathanarat}, {Janka},
  {Gabler}, {Alp}, {Diehl}, {Maeda}, {Larsson}, {Fransson}, {Menon}, \&
  {Heger}}]{jerkstrand20}
{Jerkstrand}, A., {Wongwathanarat}, A., {Janka}, H.~T., {et~al.} 2020, arXiv
  e-prints, arXiv:2003.05156

\bibitem[{{Katsuda} {et~al.}(2018){Katsuda}, {Morii}, {Janka},
  {Wongwathanarat}, {Nakamura}, {Kotake}, {Mori}, {M{\"u}ller}, {Takiwaki},
  {Tanaka}, {Tominaga}, \& {Tsunemi}}]{katsuda18}
{Katsuda}, S., {Morii}, M., {Janka}, H.-T., {et~al.} 2018, \apj, 856, 18

\bibitem[{{Lopez} {et~al.}(2019){Lopez}, {Williams}, {Safi-Harb}, {Park},
  {Plucinsky}, {Pooley}, {Temim}, {Auchettl}, {Badenes}, {Bamba}, {Castro},
  {Garofali}, {Leahy}, {Slane}, {Vink}, {Williams}, \& {Wheeler}}]{lopez19}
{Lopez}, L., {Williams}, B.~J., {Safi-Harb}, S., {et~al.} 2019, \baas, 51, 454

\bibitem[{{Lopez} {et~al.}(2009{\natexlab{a}}){Lopez}, {Ramirez-Ruiz},
  {Badenes}, {Huppenkothen}, {Jeltema}, \& {Pooley}}]{lopez09b}
{Lopez}, L.~A., {Ramirez-Ruiz}, E., {Badenes}, C., {et~al.} 2009{\natexlab{a}},
  \apjl, 706, L106

\bibitem[{{Lopez} {et~al.}(2011){Lopez}, {Ramirez-Ruiz}, {Huppenkothen},
  {Badenes}, \& {Pooley}}]{lopez11}
{Lopez}, L.~A., {Ramirez-Ruiz}, E., {Huppenkothen}, D., {Badenes}, C., \&
  {Pooley}, D.~A. 2011, \apj, 732, 114

\bibitem[{{Lopez} {et~al.}(2009{\natexlab{b}}){Lopez}, Ramirez-Ruiz, Pooley, \&
  Jeltema}]{Lopez_2009}
{Lopez}, L.~A., Ramirez-Ruiz, E., Pooley, D.~A., \& Jeltema, T.~E.
  2009{\natexlab{b}}, The Astrophysical Journal, 691, 875

\bibitem[{{Milisavljevic} \& {Fesen}(2013)}]{2013ApJ...772..134M}
{Milisavljevic}, D. \& {Fesen}, R.~A. 2013, \apj, 772, 134

\bibitem[{{M{\"u}ller}(2016)}]{muller16}
{M{\"u}ller}, B. 2016, Publications of the Astronomical Society of Australia,
  33, e048

\bibitem[{Müller {et~al.}(2019)Müller, Tauris, Heger, Banerjee, Qian, Powell,
  Chan, Gay, \& Langer}]{M_ller_2019}
Müller, B., Tauris, T.~M., Heger, A., {et~al.} 2019, Monthly Notices of the
  Royal Astronomical Society, 484, 3307–3324

\bibitem[{{Nagakura} {et~al.}(2019){Nagakura}, {Sumiyoshi}, \&
  {Yamada}}]{2019ApJ...880L..28N}
{Nagakura}, H., {Sumiyoshi}, K., \& {Yamada}, S. 2019, \apjl, 880, L28

\bibitem[{{Nigro} {et~al.}(2019){Nigro}, {Deil}, {Zanin}, {Hassan}, {King},
  {Ruiz}, {Saha}, {Terrier}, {Br{\"u}gge}, {N{\"o}the}, {Bird}, {Lin},
  {Aleksi{\'c}}, {Boisson}, {Contreras}, {Donath}, {Jouvin}, {Kelley-Hoskins},
  {Khelifi}, {Kosack}, {Rico}, \& {Sinha}}]{gammapy:2019}
{Nigro}, C., {Deil}, C., {Zanin}, R., {et~al.} 2019, \aap, 625, A10

\bibitem[{Nordhaus {et~al.}(2012)Nordhaus, Brandt, Burrows, \&
  Almgren}]{Nordhaus_2012}
Nordhaus, J., Brandt, T.~D., Burrows, A., \& Almgren, A. 2012, Monthly Notices
  of the Royal Astronomical Society, 423, 1805–1812

\bibitem[{Nordhaus {et~al.}(2010)Nordhaus, Brandt, Burrows, Livne, \&
  Ott}]{Nordhaus_2010}
Nordhaus, J., Brandt, T.~D., Burrows, A., Livne, E., \& Ott, C.~D. 2010,
  Physical Review D, 82

\bibitem[{Oliphant(2006)}]{oliphant2006guide}
Oliphant, T.~E. 2006, A guide to NumPy, Vol.~1 (Trelgol Publishing USA)

\bibitem[{{Orlando} {et~al.}(2016){Orlando}, {Miceli}, {Pumo}, \&
  {Bocchino}}]{Orlando2016}
{Orlando}, S., {Miceli}, M., {Pumo}, M.~L., \& {Bocchino}, F. 2016, \apj, 822,
  22

\bibitem[{{Patnaude} \& {Fesen}(2014)}]{Patnaude_2014}
{Patnaude}, D.~J. \& {Fesen}, R.~A. 2014, \apj, 789, 138

\bibitem[{{Picquenot} {et~al.}(2019){Picquenot}, {Acero}, {Bobin}, {Maggi},
  {Ballet}, \& {Pratt}}]{picquenot:hal-02160434}
{Picquenot}, A., {Acero}, F., {Bobin}, J., {et~al.} 2019, \aap, 627, A139

\bibitem[{{Price-Whelan} {et~al.}(2018){Price-Whelan}, {Sip{\H{o}}cz},
  {G{\"u}nther}, {Lim}, {Crawford}, {Conseil}, {Shupe}, {Craig}, {Dencheva},
  {Ginsburg}, {VanderPlas}, {Bradley}, {P{\'e}rez-Su{\'a}rez}, {de Val-Borro},
  {Paper Contributors}, {Aldcroft}, {Cruz}, {Robitaille}, {Tollerud},
  {Coordination Committee}, {Ardelean}, {Babej}, {Bach}, {Bachetti}, {Bakanov},
  {Bamford}, {Barentsen}, {Barmby}, {Baumbach}, {Berry}, {Biscani}, {Boquien},
  {Bostroem}, {Bouma}, {Brammer}, {Bray}, {Breytenbach}, {Buddelmeijer},
  {Burke}, {Calderone}, {Cano Rodr{\'\i}guez}, {Cara}, {Cardoso}, {Cheedella},
  {Copin}, {Corrales}, {Crichton}, {D{\textquoteright}Avella}, {Deil},
  {Depagne}, {Dietrich}, {Donath}, {Droettboom}, {Earl}, {Erben}, {Fabbro},
  {Ferreira}, {Finethy}, {Fox}, {Garrison}, {Gibbons}, {Goldstein}, {Gommers},
  {Greco}, {Greenfield}, {Groener}, {Grollier}, {Hagen}, {Hirst}, {Homeier},
  {Horton}, {Hosseinzadeh}, {Hu}, {Hunkeler}, {Ivezi{\'c}}, {Jain}, {Jenness},
  {Kanarek}, {Kendrew}, {Kern}, {Kerzendorf}, {Khvalko}, {King}, {Kirkby},
  {Kulkarni}, {Kumar}, {Lee}, {Lenz}, {Littlefair}, {Ma}, {Macleod},
  {Mastropietro}, {McCully}, {Montagnac}, {Morris}, {Mueller}, {Mumford},
  {Muna}, {Murphy}, {Nelson}, {Nguyen}, {Ninan}, {N{\"o}the}, {Ogaz}, {Oh},
  {Parejko}, {Parley}, {Pascual}, {Patil}, {Patil}, {Plunkett}, {Prochaska},
  {Rastogi}, {Reddy Janga}, {Sabater}, {Sakurikar}, {Seifert}, {Sherbert},
  {Sherwood-Taylor}, {Shih}, {Sick}, {Silbiger}, {Singanamalla}, {Singer},
  {Sladen}, {Sooley}, {Sornarajah}, {Streicher}, {Teuben}, {Thomas},
  {Tremblay}, {Turner}, {Terr{\'o}n}, {van Kerkwijk}, {de la Vega}, {Watkins},
  {Weaver}, {Whitmore}, {Woillez}, {Zabalza}, \& {Contributors}}]{astropy:2018}
{Price-Whelan}, A.~M., {Sip{\H{o}}cz}, B.~M., {G{\"u}nther}, H.~M., {et~al.}
  2018, \aj, 156, 123

\bibitem[{{Reed} {et~al.}(1995){Reed}, {Hester}, {Fabian}, \&
  {Winkler}}]{1995ApJ...440..706R}
{Reed}, J.~E., {Hester}, J.~J., {Fabian}, A.~C., \& {Winkler}, P.~F. 1995,
  \apj, 440, 706

\bibitem[{{Schure} {et~al.}(2008){Schure}, {Vink}, {Garc{\'\i}a-Segura}, \&
  {Achterberg}}]{schure08}
{Schure}, K.~M., {Vink}, J., {Garc{\'\i}a-Segura}, G., \& {Achterberg}, A.
  2008, \apj, 686, 399

\bibitem[{{Summa} {et~al.}(2018){Summa}, {Janka}, {Melson}, \&
  {Marek}}]{summa18}
{Summa}, A., {Janka}, H.-T., {Melson}, T., \& {Marek}, A. 2018, \apj, 852, 28

\bibitem[{{Tashiro} {et~al.}(2018){Tashiro}, {Maejima}, {Toda}, {Kelley},
  {Reichenthal}, {Lobell}, {Petre}, {Guainazzi}, {Costantini}, {Edison},
  {Fujimoto}, {Grim}, {Hayashida}, {den Herder}, {Ishisaki}, {Paltani},
  {Matsushita}, {Mori}, {Sneiderman}, {Takei}, {Terada}, {Tomida}, {Akamatsu},
  {Angelini}, {Arai}, {Awaki}, {Babyk}, {Bamba}, {Barfknecht}, {Barnstable},
  {Bialas}, {Blagojevic}, {Bonafede}, {Brambora}, {Brenneman}, {Brown},
  {Brown}, {Burns}, {Canavan}, {Carnahan}, {Chiao}, {Comber}, {Corrales}, {de
  Vries}, {Dercksen}, {Diaz-Trigo}, {Dillard}, {DiPirro}, {Done}, {Dotani},
  {Ebisawa}, {Eckart}, {Enoto}, {Ezoe}, {Ferrigno}, {Fukazawa}, {Fujita},
  {Furuzawa}, {Gallo}, {Graham}, {Gu}, {Hagino}, {Hamaguchi}, {Hatsukade},
  {Hawes}, {Hayashi}, {Hegarty}, {Hell}, {Hiraga}, {Hodges-Kluck}, {Holland},
  {Hornschemeier}, {Hoshino}, {Ichinohe}, {Iizuka}, {Ishibashi}, {Ishida},
  {Ishikawa}, {Ishimura}, {James}, {Kallman}, {Kara}, {Katsuda}, {Kenyon},
  {Kilbourne}, {Kimball}, {Kitaguti}, {Kitamoto}, {Kobayashi}, {Kohmura},
  {Koyama}, {Kubota}, {Leutenegger}, {Lockard}, {Loewenstein}, {Maeda},
  {Marbley}, {Markevitch}, {Matsumoto}, {Matsuzaki}, {McCammon}, {McNamara},
  {Miko}, {Miller}, {Miller}, {Minesugi}, {Mitsuishi}, {Mizuno}, {Mori},
  {Mukai}, {Murakami}, {Mushotzky}, {Nakajima}, {Nakamura}, {Nakashima},
  {Nakazawa}, {Natsukari}, {Nigo}, {Nishioka}, {Nobukawa}, {Nobukawa}, {Noda},
  {Odaka}, {Ogawa}, {Ohashi}, {Ohno}, {Ohta}, {Okajima}, {Okamoto}, {Onizuka},
  {Ota}, {Ozaki}, {Plucinsky}, {Porter}, {Pottschmidt}, {Sato}, {Sato},
  {Sawada}, {Seta}, {Shelton}, {Shibano}, {Shida}, {Shidatsu}, {Shirron},
  {Simionescu}, {Smith}, {Someya}, {Soong}, {Suagawara}, {Szymkowiak},
  {Takahashi}, {Tamagawa}, {Tamura}, {Tanaka}, {Terashima}, {Tsuboi},
  {Tsujimoto}, {Tsunemi}, {Tsuru}, {Uchida}, {Uchiyama}, {Ueda}, {Uno},
  {Walsh}, {Watanabe}, {Williams}, {Wolfs}, {Wright}, {Yamada}, {Yamaguchi},
  {Yamaoka}, {Yamasaki}, {Yamauchi}, {Yamauchi}, {Yanagase}, {Yaqoob},
  {Yasuda}, {Yoshioka}, {Zabala}, \& {Irina}}]{tashiro18}
{Tashiro}, M., {Maejima}, H., {Toda}, K., {et~al.} 2018, in Society of
  Photo-Optical Instrumentation Engineers (SPIE) Conference Series, Vol. 10699,
  Space Telescopes and Instrumentation 2018: Ultraviolet to Gamma Ray, 1069922

\bibitem[{{Thorstensen} {et~al.}(2001){Thorstensen}, {Fesen}, \& {van den
  Bergh}}]{thorstensen01}
{Thorstensen}, J.~R., {Fesen}, R.~A., \& {van den Bergh}, S. 2001, \aj, 122,
  297

\bibitem[{{Williams} {et~al.}(2019){Williams}, {Auchettl}, {Badenes}, {Castro},
  {Kargaltsev}, {Lopez}, {Mori}, {Patnaude}, {Plucinsky}, {Raymond},
  {Safi-Harb}, {Slane}, {Tanaka}, {Temim}, {Vink}, \& {Yamaguchi}}]{williams19}
{Williams}, B., {Auchettl}, K., {Badenes}, C., {et~al.} 2019, \baas, 51, 263

\bibitem[{{Willingale} {et~al.}(2002{\natexlab{a}}){Willingale}, {Bleeker},
  {van der Heyden}, {Kaastra}, \& {Vink}}]{2002A&A...381.1039W}
{Willingale}, R., {Bleeker}, J.~A.~M., {van der Heyden}, K.~J., {Kaastra},
  J.~S., \& {Vink}, J. 2002{\natexlab{a}}, \aap, 381, 1039

\bibitem[{{Willingale} {et~al.}(2002{\natexlab{b}}){Willingale}, {Bleeker},
  {van der Heyden}, {Kaastra}, \& {Vink}}]{willingale02}
{Willingale}, R., {Bleeker}, J.~A.~M., {van der Heyden}, K.~J., {Kaastra},
  J.~S., \& {Vink}, J. 2002{\natexlab{b}}, \aap, 381, 1039

\bibitem[{{Wongwathanarat} {et~al.}(2013){Wongwathanarat}, {Janka}, \&
  {M{\"u}ller}}]{wongwathanarat13}
{Wongwathanarat}, A., {Janka}, H.-T., \& {M{\"u}ller}, E. 2013, \aap, 552, A126

\bibitem[{{Wongwathanarat} {et~al.}(2017){Wongwathanarat}, {Janka},
  {M{\"u}ller}, {Pllumbi}, \& {Wanajo}}]{wongwathanarat17}
{Wongwathanarat}, A., {Janka}, H.-T., {M{\"u}ller}, E., {Pllumbi}, E., \&
  {Wanajo}, S. 2017, \apj, 842, 13

\end{thebibliography}

\begin{appendix} %First appendix
\section{Ionization impact on line centroid}
\label{ap:velocity}
\begin{figure}
\centering
\includegraphics[width=7cm]
{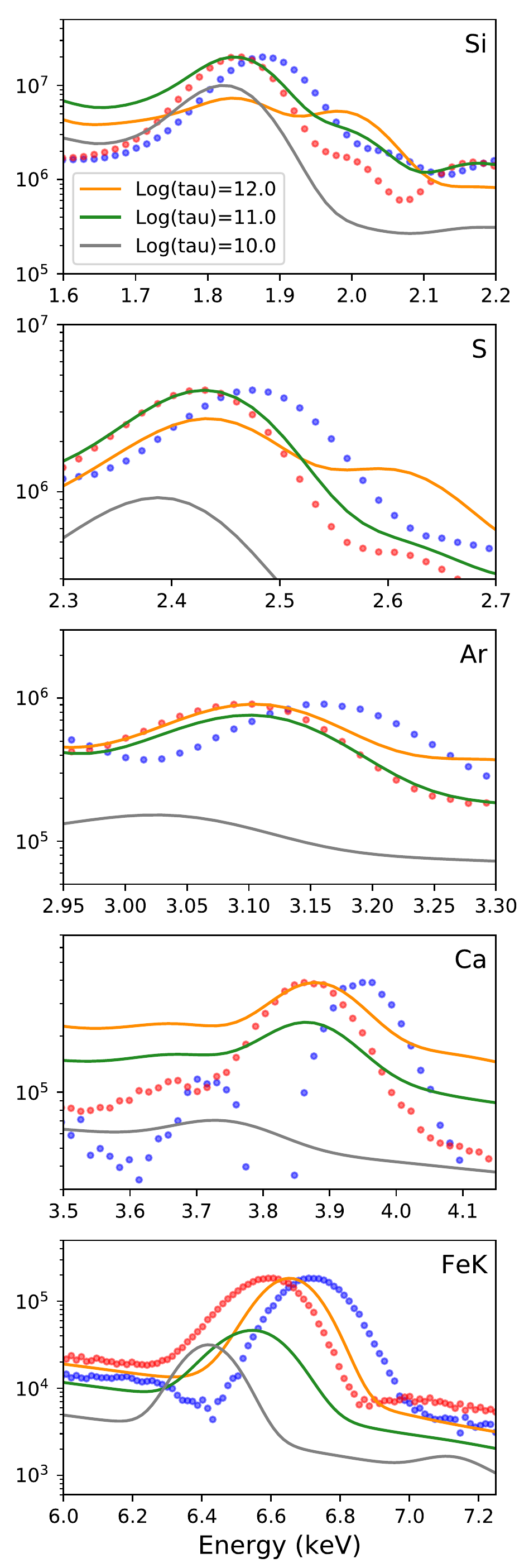}
\caption{Comparison of our red and blue spectra (dotted curves) presented in Fig.~\ref{fig:red_blue_images} versus $pshock$ Xspec models with different ionization timescales for  $kT$=1.5 keV. The y-axis is in counts.}
\label{fig:xspec_model}
\end{figure}

At the spectral resolution of CCD type instruments, most emission lines are not resolved and the observed emission is a blurred complex of lines.
The centroid energy of emission lines can shift either via Doppler effect or when the ionization timescale increases and the ions distribution in a given line complex evolves \textbf{\citep{dewey2010,greco20}}.
In Fig.~\ref{fig:xspec_model} we compare the spectral model $pshock$ at different ionization timescales with the spectra that we labeled red- and blue-shifted in Fig.~\ref{fig:red_blue_images}. The temperature of the model was fixed to 1.5 keV based on the temperature histogram of Fig.~2 of \citet{hwang12}. The effective area and redistribution matrix from observation ObsID: 4634 were used.
We can see that as the ionization timescale $\tau$ increases the line centroid, which is a blend of multiple lines, shifts to higher energies. This is most visible in the Fe-K region where a large number of lines exists.
We note that the spectral component that we labeled as blue-shifted is well beyond any ionization state shown here and reinforces the idea that this component is dominated by velocity effect.
The situation is less clear for the red-shifted component where the
shift in energy is not as strong. We also do not precisely know which
reference line it can be compared to. 
It is interesting to note that for the purpose of measuring a velocity effect while minimizing the confusion with ionization effects, the Ar and Ca lines provide the best probe. Indeed for $\tau > 10^{11}$ cm$^{-3}$ s, the centroid of the main Ar and Ca lines shows no evolution given the CCD energy resolution.
The Fe K centroid strongly varies with $\tau$ and the choice of a reference ionization state and reference energy limits the reliability of this line for velocity measurements in non-equilibrium ionization plasma.

\section{Retrieving error bars for a nonlinear estimator applied on a Poissonian data set}
\label{ap:boot}
\subsection{Introduction}

    The BSS method we used in this paper, the pGMCA, is one of the numerous advanced data analysis methods that have recently been introduced for a use in astrophysics, among which we can also find other BSS methods, classification, PSF deconvolution, denoising or dimensionality reduction. We can formalize the application of these data analysis methods by writing $\Theta=A(X)$ , where $X$ is the original data, $A$ is the nonlinear analysis operator used to process the signal and $\Theta$ is the estimator for which we want to find errors (in this paper for example, $X$ is the original X-ray data from Cas A, $A$ is the pGMCA algorithm and $\Theta$ represents the retrieved spectra and images). Most of these methods being nonlinear, there is no easy way to retrieve error bars or a confidence interval associated with the estimator $\Theta$. Estimating errors accurately in a nonlinear problem is still an open question that goes far beyond the scope of astrophysical applications as there is no general method to get error bars from a nonlinear data-driven method such as the pGMCA. This is a hot topic whose study would be essential for an appropriate use of complex data analysis methods in retrieving physical parameters, and for allowing the user to estimate the accuracy of the results.
    
    \subsection{Existing methods to retrieve error bars on Poissonian data sets}
    
    Our aim, when searching for error bars associated with a certain estimator $\Theta$ on an analyzed data set, is to obtain the variance of $\Theta=A(X)$ , where the original data $X$ is composed of $N$ elements. When working on a simulation, an obvious way to proceed in order to estimate the variance of $\Theta$ is to apply the considered data analysis method $A$ on a certain number of Monte-Carlo (MC) realizations $X_i$ and look at the standard deviation of the results $\Theta_i=A({X_i})$. The variance of the $\Theta_i$ provides a good estimation of the errors. Yet, this cannot be done with real data as only one observation is available: the observed one. Thus, a resampling method such as the jackknife, the bootstrap \citep[see][]{befron79} or its derivatives, able to simulate several realizations out of a single one, is necessary. Ideally, the aim is to obtain through this resampling method a number of ''fake'' MC realizations centered on the original data: new data sets variating spatially and reproducing the spread of MC drawings with a mean equal or close to the mean of the original data. 
    
    The mechanisms at stake in jacknife or bootstrap resamplings are similar. Jacknife and bootstrap resampling methods produce $n$ resampled sets ${\tilde{X_i}}$ by rearranging the elements of $X$, and allow us to consider the variance of $\Theta_i=A({\tilde{X_i}})$ for $i$ in $ \llbracket 1, n \rrbracket$ as an approximation for the variance of $\Theta$. As jacknife and bootstrap methods are close to each other, and the bootstrap and some of its derivatives are more adapted to handle correlated data sets, we will in this Appendix focus on a particular method, representative of other resampling methods and theoretically suited for astrophysical applications: the block bootstrap, which is a simple bootstrap applied on randomly formed groups of events rather than on the individual events.
  
\begin{figure}
\centering
\includegraphics[width = 6cm]{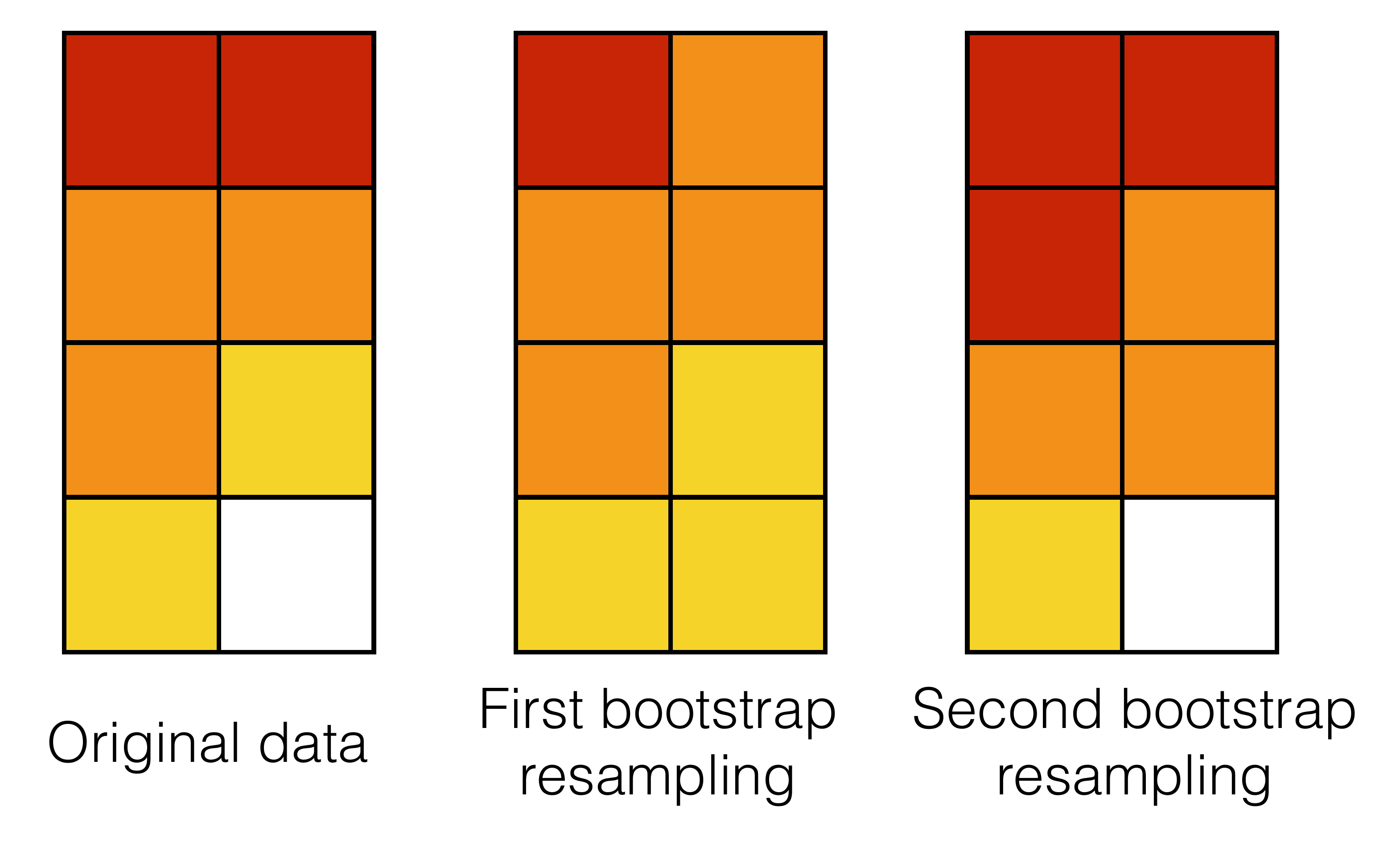}
\caption{Example of bootstrap resampling. Each square represents a different event, each color a different value. $N$ events are taken randomly with replacement from the original data to create each of the two bootstrap resamplings.}
\label{fig:resamples}
\end{figure}

    In the case of a Poisson process, the discrete nature of the elements composing the data set can easily be resampled with a block bootstrap method. The $N$ discrete elements composing a Poissonian data set $X$ will be called "events." In X-rays for example, the events are the photons detected by the spectro-imaging instrument. The bootstrap consists in a random sampling with replacement from the current set events $X$. The resampling obtained through bootstrapping is a set $\tilde{X}^{boot}$ of $N$ events taken randomly with replacement amid the initial ones (see Fig.~\ref{fig:resamples}). This method can be repeated in order to simulate as many realizations $\tilde{X}^{boot}_i$ as needed to estimate standard errors or confidence intervals. In order to save calculation time, we can choose to resample blocks of data of a fixed size instead of single events: This method is named block bootstrap. The block bootstrap is also supposed to conserve correlations more accurately, making it more appropriate for a use on astrophysical signals. The data can be of any dimension but for clarity, we will only show in this Appendix bi-dimensional data sets, that is images. 

\subsection{Biases in classical bootstrap applied on Poissonian data sets}
\label{sec:boot}

	The properties of the data resampled strongly depends on the nature of the original data. Biases may appear in the resampled data sets, proving a block bootstrap can fail to reproduce consistent data that could be successfully used to evaluate the accuracy of certain estimators.
	
	In particular, Poissonian data sets, including our X-rays data of Cas~A, are not consistently resampled by current resampling methods such as the block bootstrap. A Poissonian data set $X$ can be defined as a Poisson realization of an underlying theoretical model $X^*$, which can be written: 
	\begin{equation*}
	X=\mathcal{P}(X^*),    
	\end{equation*}
	where $\mathcal{P}(.)$ is an operator giving as an output a Poisson realization of a set. 
	
\begin{figure}
\centering
\subfloat{\includegraphics[width = 4cm]{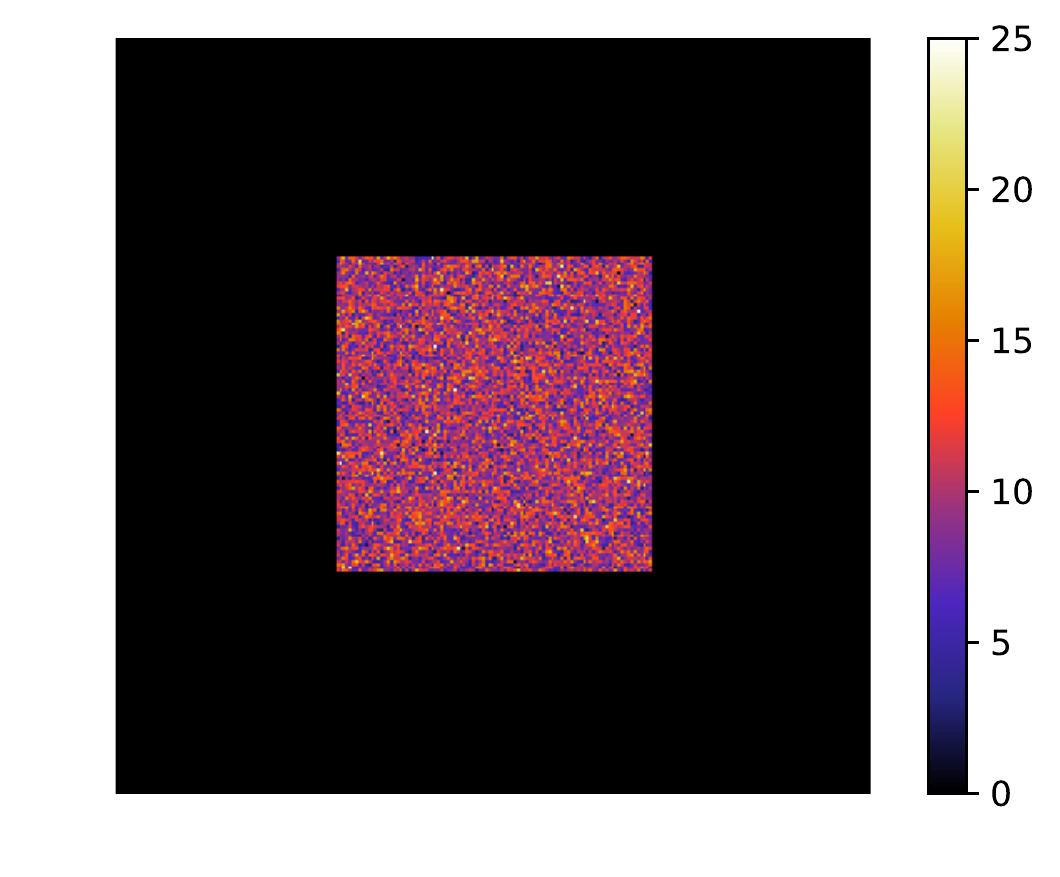}}
\subfloat{\includegraphics[width = 5.cm]{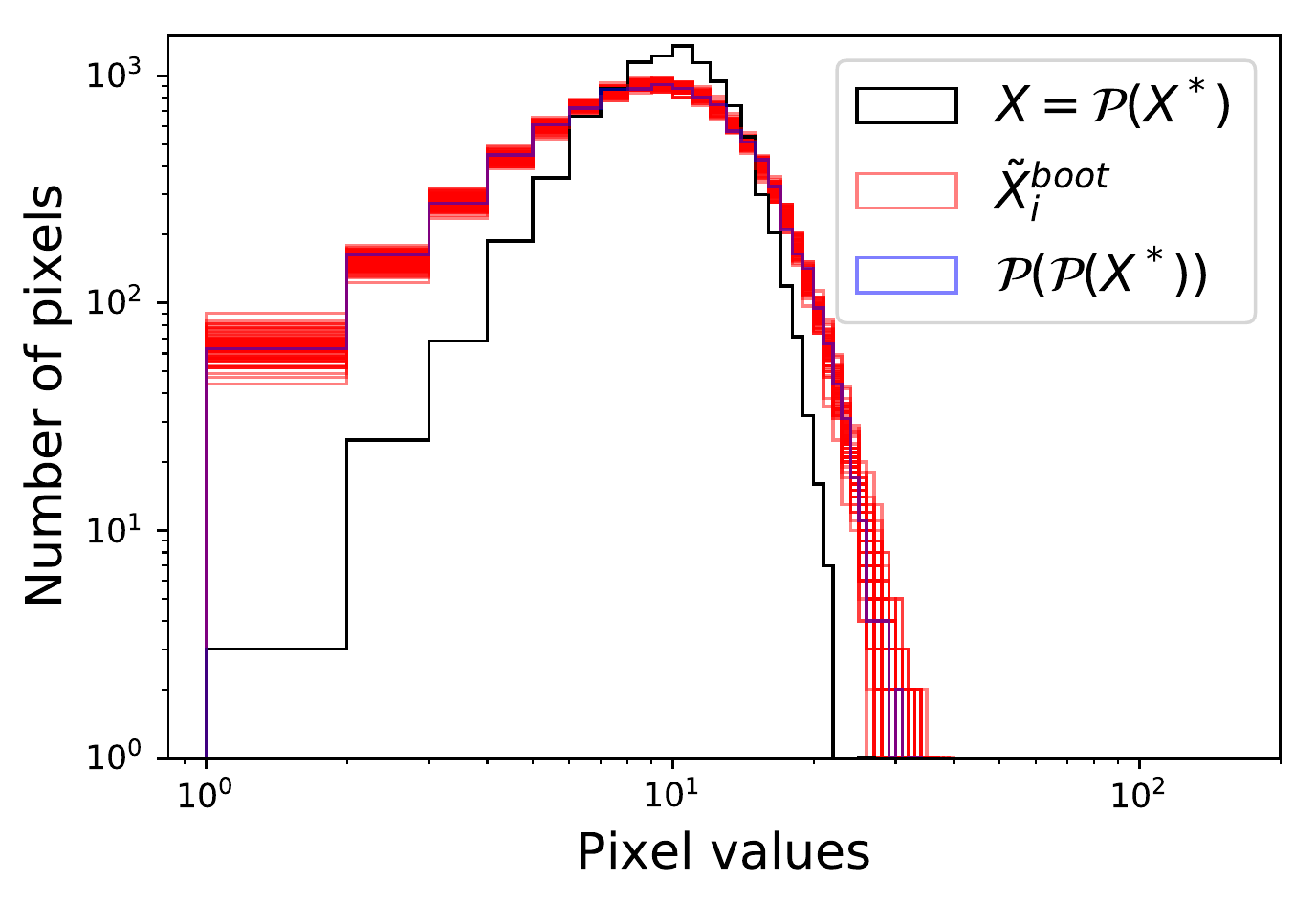}}\\
\subfloat{\includegraphics[width = 4cm]{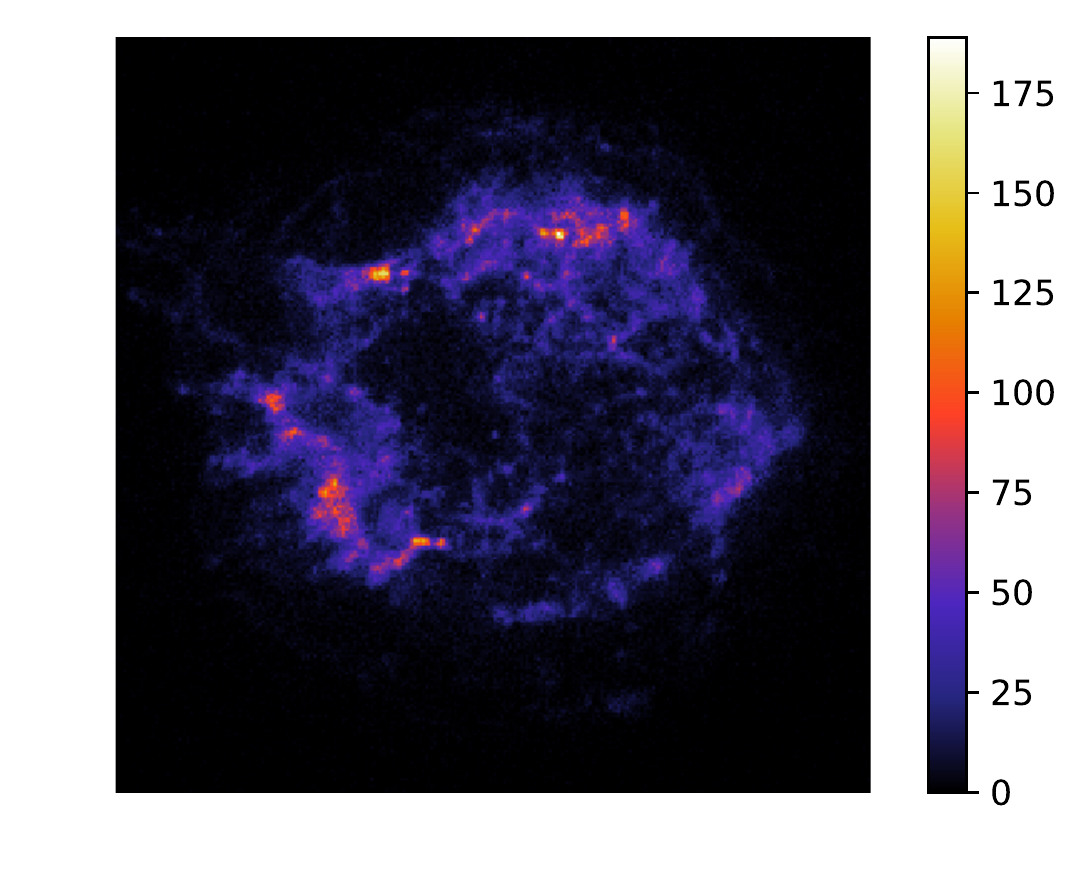}}
\subfloat{\includegraphics[width = 5.cm]{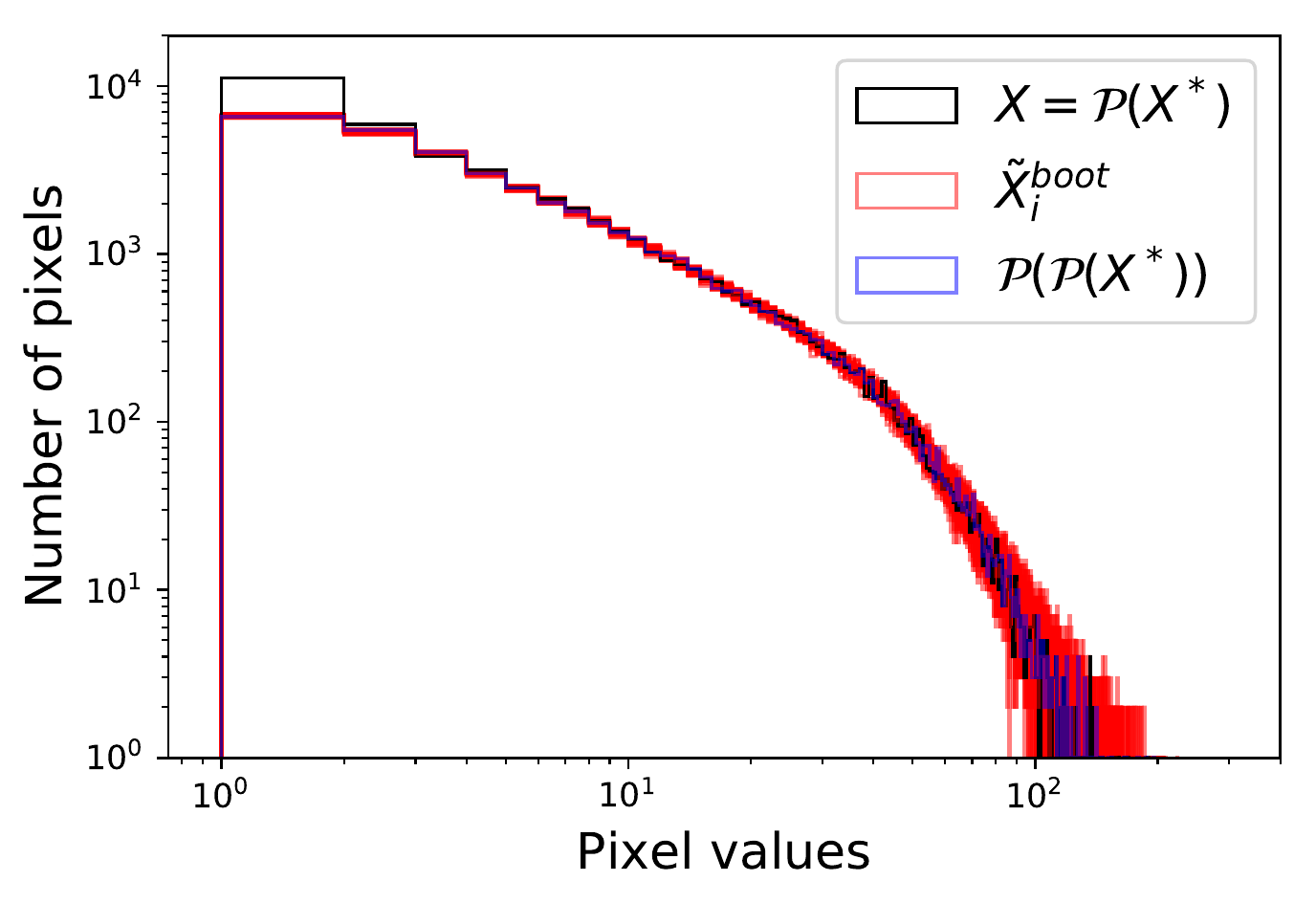}}\\
\caption{Data sets and their associated histogram in two cases: on top, the very simple case of a Poisson realization of the image of a square with uniform value $10$; on the bottom, a toy model Cas~A image obtained by taking a Poisson realization of a high-statistics denoised image of Cas~A (hereafter called toy model). On the right, the black histogram correspond to the original data $X=\mathcal{P}(X^*)$. The red histograms are those of the data sets $\tilde{X}^{boot}_i$ obtained through resampling of the original data and the blue ones are the histograms of a Poisson realization of the original data $\mathcal{P}(X)=\mathcal{P}(\mathcal{P}(X^*))$. It appears that the resampled data sets have histograms highly similar to that of the original data with additional Poisson noise.}
\vspace{-0.5cm}
\label{fig:histo_pix}
\end{figure}

	A look on the histogram of a data set resampled from a Poissonian signal shows the block bootstrap fails to reproduce accurately the characteristics of the original data. Fig.~\ref{fig:histo_pix}, top, compares the histogram of the real data $X$, a simple image of a square with Poisson noise, with the histograms of the resampled data sets $\tilde{X}^{boot}_i$, and highlights the fact that the latter are more similar to the histogram of a Poisson realization of the original data $\mathcal{P}(X)=\mathcal{P}(\mathcal{P}(X^*))$ than to the actual histogram of the original data $X=\mathcal{P}(X^*)$, where $X^*$ is the underlying model of a square before adding Poisson noise. This is consistent with the fact that the block bootstrap is a random sampling with replacement, which introduces uncertainties of the same nature as a Poisson drawing.
	
\begin{figure}
\includegraphics[width = 9cm]{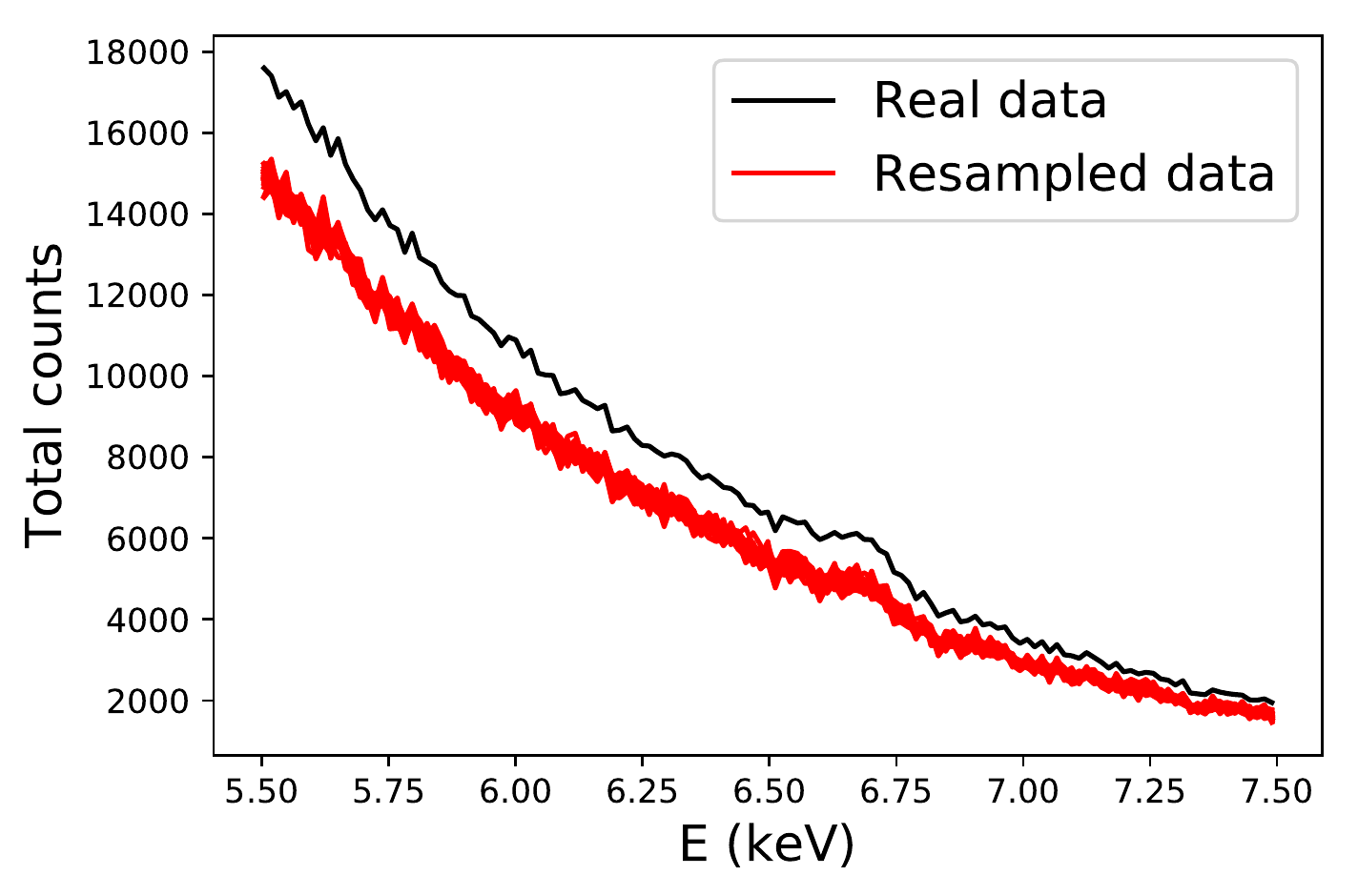}
\caption{Spectrum of the synchrotron component retrieved by pGMCA on the $5.5$-$7.5$ keV energy band on real data and on a set of 30 block bootstrap resamples. There is an obvious bias in the results, the resampled data spectra being consistently underestimated.}
\label{fig:bias_sync}
\end{figure}

	Fig.~\ref{fig:histo_pix}, bottom, shows the comparison between the histogram of the toy model Cas~A image and the histograms of the data sets resampled with a block bootstrap. We can see the resampling is, in this case too, adding Poissonian noise and gives histograms resembling $\mathcal{P}(\mathcal{P}(X^*))$ rather than $\mathcal{P}(X^*)$. The same goes with our real data cube of Cas~A: Fig.~\ref{fig:bias_sync} shows an obvious instance of this bias being transferred to the results of the pGMCA, thus proving the block bootstrap cannot be used as such to retrieve error bars for this algorithm.

\subsection{A new constrained bootstrap method}
\label{sec:newboot}

Bootstrap resamplings consisting in random drawings with replacement, it is natural that they fail to reproduce some characteristics of the data, among which the histogram that gets closer to the histogram of a Poisson realization of the original data than to the histogram of the actual data. The block bootstrap method is therefore unable to simulate a MC centered on the original data: the alteration of the histogram strongly impacts the nature of the data, hence the differences in the morphologies observed by looking at the wavelet coefficients. It is then necessary to find a new method in which we could force the histogram of the resampled data sets to be similar to that of the original data.

\begin{figure}
\centering
\subfloat{\includegraphics[width = 9cm]{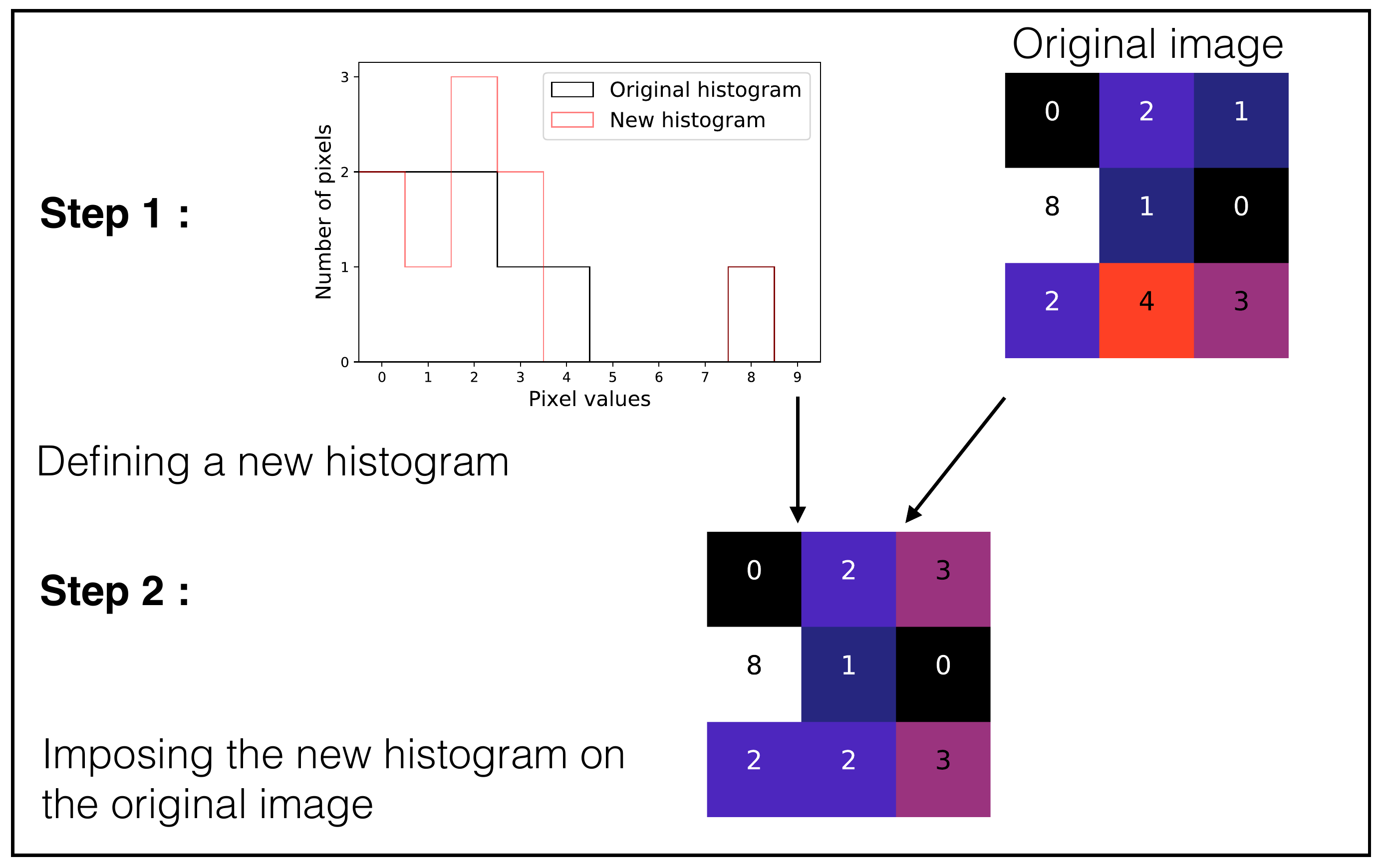}}
\\
\caption{Scheme resuming the two steps of our new constrained bootstrap method.}
\label{fig:scheme}
\end{figure}

A natural way to do so would be to impose the histogram we want the resampled data to have before actually resampling the data. To allow this constraint to be made on the pixel distribution, we can no longer consider our events to be the individual elements of $X$ or a block assembling a random sample of them. We should directly work on the pixels and their values, the pixels here being the basic bricks constituting our data. Just as the block bootstrap, our new method can work with data of any dimension. In the case of images, the "basic bricks" correspond to actual pixels values. In the case of X-rays data cubes, they are tiny cubes of the size of a pixel along the spatial dimensions, and the size of an energy bin along the spectral dimension. The same goes for any dimension of our original data. The method can also be adapted for uni-dimensional data sets. \textbf {The key of our new method is then to work on the histogram of the data presenting the pixels' values rather than on the data itself, event by event.}

We can either change the value of a pixel or exchange the value of a pixel for that of another one. The first operation simultaneously adds and subtracts $1$ in the corresponding columns of the global histogram while the second operation does not produce any change in it. A good mixture of these two operations would then allow us to obtain the histogram we want to impose in our resampled data sets, and following a Poisson probability law to select the pixels to exchange would introduce some spatial variations, in order to reproduce what a MC would do.

Our new constrained bootstrap method is thus composed of two steps, that are described below and illustrated in Fig.~\ref{fig:scheme}:

Firstly, obtaining the probability density function of the random variable underlying the observed data histogram using the Kernel Density Estimation (KDE), and randomly generating $n$ histograms from this density function with a spread around the data mimicking that of a MC, with a constraint enforcing a Poissonian distribution of the total sums of pixel values of the $n$ histograms.

Secondly, producing resampled data sets associated with the new histograms by changing the values of wisely chosen pixels in the original image.

During these steps, the pixels equal to zero remain equal to zero, and the nonzero pixels keep a strictly positive value. This constraint enforces the number of nonzero pixels to be constant and avoids the creation of random emergence of nonzero pixels in the empty area of the original data. While this is not completely realistic we prefer constraining the resampled data sets in this way than getting spurious features. We could explore ways to release this constraint in the future.

\begin{figure}
\centering
\subfloat{\includegraphics[width = 4.5cm]{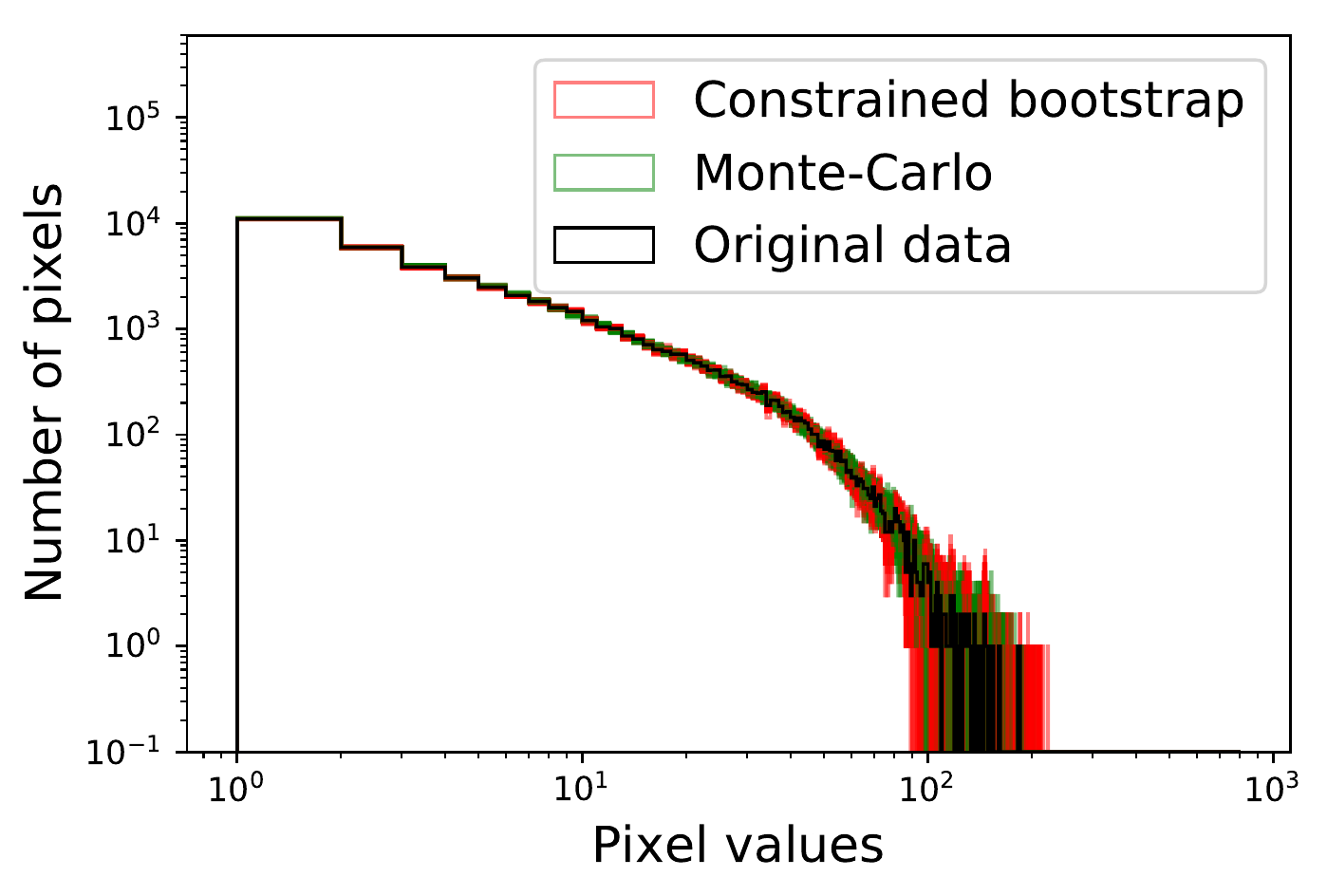}}
\subfloat{\includegraphics[width = 4.5cm]{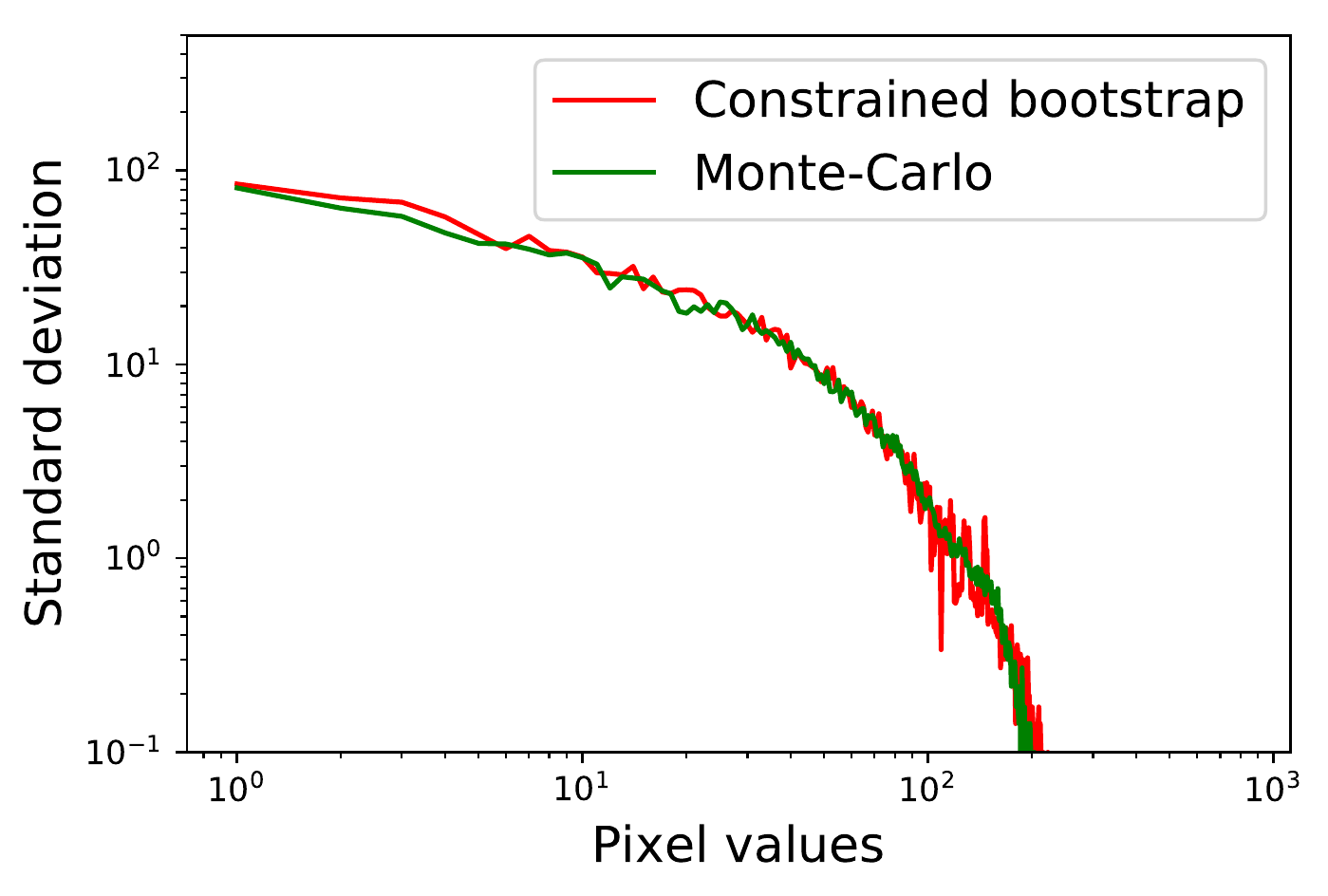}}\\

\caption{ Histograms and standard deviations of the original and resampled data sets. On the left, histograms of the original data, the resampled data sets and the MC realizations of the toy model Cas~A image. On the right, the standard deviations of the resampled data sets and MC realizations bin by bin of the histogram on the left. We can notice the great adequation between the standard deviations of the resampled data sets and that of the MC realizations.}

\label{fig:histo_comp_MC}
\end{figure}

\begin{figure}
\includegraphics[width = 9cm]{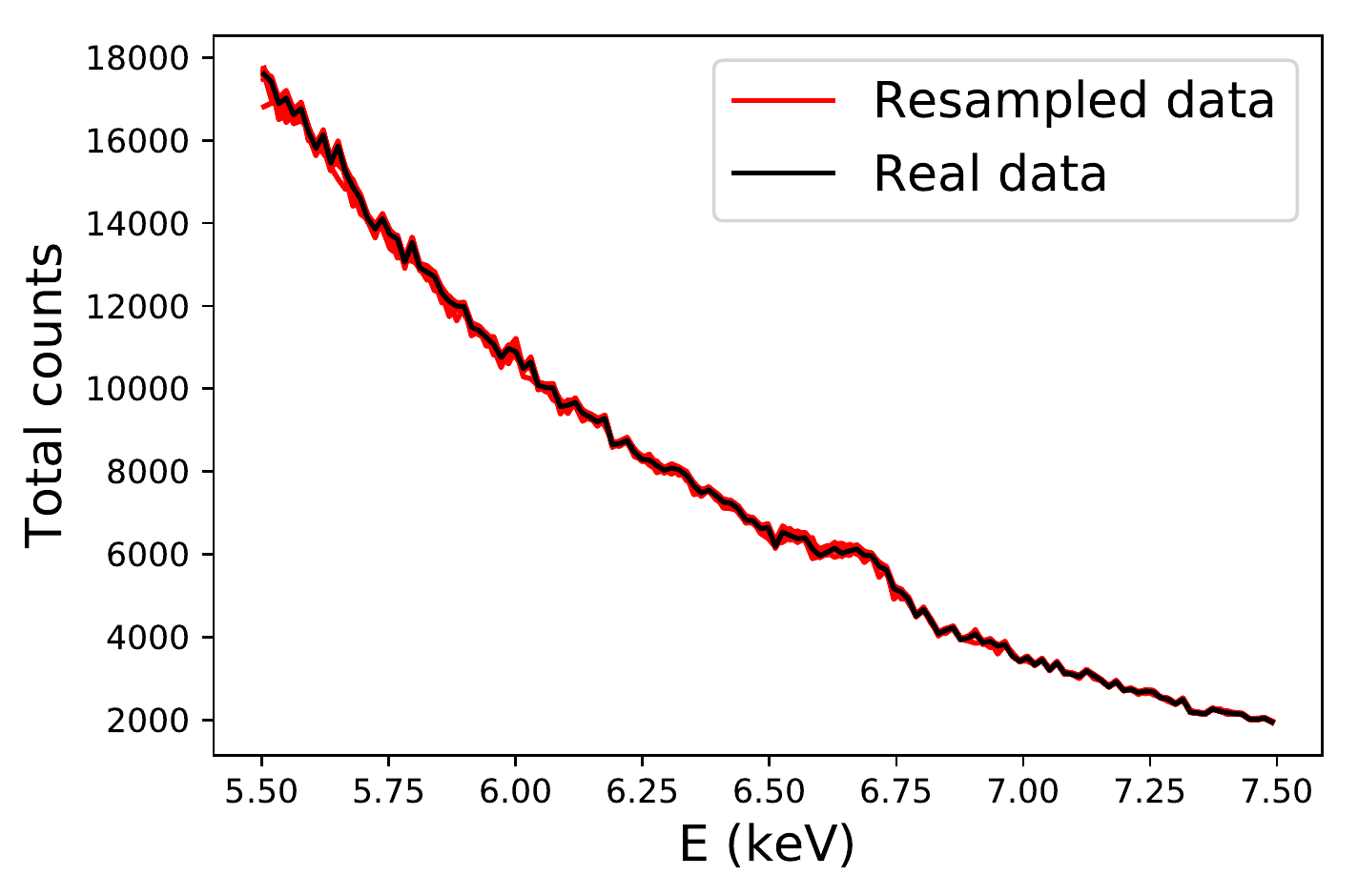}
\caption{Spectrum of the synchrotron component retrieved by pGMCA on the $5.5$-$7.5$ keV energy band on real data and on a set of 100 constrained bootstrap resamples. The bias we observed in Fig.~\ref{fig:bias_sync} between the real Cas A data and its block bootstrap resamples has been suppressed with our new constrained bootstrap method.}
\label{fig:nobias_sync}
\end{figure}

Fig.~\ref{fig:histo_comp_MC} highlights the similarities between the original histogram and those obtained through MC realizations and our new constrained bootstrap resamplings, while Fig.~\ref{fig:nobias_sync} and the spectra in Fig.~\ref{fig:red_blue_images} and Fig.~\ref{fig:other_images} show that even after being processed by the sensitive pGMCA algorithm, this resampling method shows little to no biases. Hence, our new constrained bootstrap method brings a first and successful attempt at solving the problem of biases in bootstrapping Poissonian data sets. 

The comparison of our resampled data sets to a group of MC realizations of the same simulation of Cas~A appears to be promising for the variance induced by our method. However, when applying a complex estimator such as the pGMCA on both the MC realizations and our resampled data sets, it appears that the variances obtained through our method fail to accurately reproduce those of the MC realizations. For that reason, the error bars retrieved by our constrained bootstrap method do not have a physical signification. Nevertheless, they constitute an interesting way to assess the robustness of our method around a certain line emission. The different resamplings explore initial conditions slightly different from the original data, thus evaluating the dependence of our results on the initial conditions. Fig.~\ref{fig:red_blue_images} and Fig.~\ref{fig:other_images} indeed show that for some line emissions, the dispersion between the results on different resampled data sets is far greater than for others.

This new constrained bootstrap method is a first and promising attempt at retrieving error bars for nonlinear estimators on Poissonian data sets, a problem that is often not trivial. In nonlinear processes, errors frequently cannot be propagated correctly, so the calculation of sensitive parameters and the estimation of errors after an extensive use of an advanced data analysis could benefit from this method. We will work in the future on a way to constraint the variance of the results to be more closely related to that of a set of MC realizations in order to ensure the physical signification of the obtained error bars.

\end{appendix}

\end{document}